\documentstyle[epsf,12pt]{article}
\begin{document}\thispagestyle{empty}\begin{flushright}
OUT--4102--46\\LTH--360\\MZ--TH/95--28\\hep-th/9607174\\
23 July 1996
            \end{flushright} \vspace*{2mm} \begin{center} {\large\bf
Beyond the triangle and uniqueness relations:\\
non-zeta counterterms at large $N$ from positive knots$^{*)}$
            } \vglue5mm{\bf
D.~J.~Broadhurst$^{1)}$
            }\vglue 2mm
Physics Department, Open University, Milton Keynes MK7 6AA, UK
            \vglue 4mm{\bf
J.A.\ Gracey$^{2)}$
            } \vglue 2mm
Department of Mathematical Sciences, University of Liverpool,\\
PO Box 147, Liverpool L69 3BX, UK
            \vglue 4mm{\bf
D.~Kreimer$^{3)}$
            } \vglue2mm
Institut f\"{u}r Physik der Universit\"{a}t Mainz, D-55099 Mainz, FRG
            \end{center} \vfill
{\small{\bf Abstract}
Counterterms that are not reducible to $\zeta_{n}$ are generated by ${}_3F_2$
hypergeometric series arising from diagrams for which triangle and uniqueness
relations furnish insufficient data. Irreducible double sums, corresponding to
the torus knots $(4,3)=8_{19}$ and $(5,3)=10_{124}$, are found in anomalous
dimensions at ${\rm O}(1/N^3)$ in the large-$N$ limit, which we compute
analytically up to terms of level 11, corresponding to 11 loops for
4-dimensional field theories and 12 loops for 2-dimensional theories.
High-precision numerical results are obtained up to 24 loops and used in Pad\'e
resummations of $\varepsilon$-expansions, which are compared with analytical
results in 3 dimensions. The ${\rm O}(1/N^3)$ results entail knots generated by
three dressed propagators in the master two-loop two-point diagram. At higher
orders in $1/N$ one encounters the uniquely positive hyperbolic 11-crossing
knot, associated with an irreducible triple sum. At 12 crossings, a pair of
3-braid knots is generated, corresponding to a pair of irreducible double sums
with alternating signs. The hyperbolic positive knots $10_{139}$ and $10_{152}$
are not generated by such self-energy insertions.
                                                    }\vfill
                                                    \footnoterule\noindent
$^*$) Work supported in part by grant CHRX--CT94--0579, from HUCAM.\\
$^1$) D.Broadhurst@open.ac.uk\\
$^2$) jag@amtp.liv.ac.uk, PPARC advanced fellow\\
$^3$) kreimer@dipmza.physik.uni-mainz.de
\newpage
\newcommand{\fk}[4]{\sigma_1^{#1}\sigma_2^{#2}\sigma_1^{#3}\sigma_2^{#4}}
\newcommand{\df}[2]{\mbox{$\frac{#1}{#2}$}}
\newcommand{\Eq}[1]{~(\ref{#1})}
\newcommand{\Eqq}[2]{~(\ref{#1},\ref{#2})}
\newcommand{\Eqqq}[3]{~(\ref{#1},\ref{#2},\ref{#3})}
\newcommand{\ep}{\varepsilon}
\newcommand{\ze}[1]{\zeta_{#1}}
\newcommand{\al}[1]{\alpha_{#1}}
\newcommand{\ba}[1]{\overline{\alpha}_{#1}}
\newcommand{\rd}{{\rm d}}
\newcommand{\hyp}{{}_3F_2}
\newcommand{\pz}[1]{\psi_{#1}}
\newcommand{\pp}[1]{\psi_{#1}^\prime}
\newcommand{\pg}[1]{\psi_{#1}^{\prime\prime}}
\newcommand{\eltft}
{\sigma_1^2\sigma_2^2\sigma_1^{}\sigma_3^{}\sigma_2^3\sigma_3^2}
\newcommand{\tline}[6]{\\#1&#2&#3&#4&#5&#6}
\newcommand{\begtab}{\begin{equation}\begin{array}{rrrrrr}%
L&[L\backslash0]&[L\!-\!2\backslash2]&[L\!-\!4\backslash4]&%
[L\!-\!6\backslash6]&[L\!-\!8\backslash8]}
\newcommand{\tabend}[1]{\end{array}\label{#1}\end{equation}}
\newcommand{\disp}[4]{\left\{\begin{array}{c}%
#1,#2\\#3,#4\end{array}\right\}}

\section{Introduction}

Triangle~\cite{CT} and uniqueness~\cite{UNI} relations
make the {\em analytical\/} work of renormalizing a field theory:
{\em elementary\/} to 3 loops~\cite{TVZ};
subject to a comprehensive {\em algorithm\/}~\cite{CT}
at 4 loops~\cite{4LQ}; and {\em achievable\/}, with ingenuity~\cite{KAZ},
in the case of $\phi^4$-theory at 5 loops~\cite{PHI}.
Then an obstacle~\cite{5LB} occurs.
Fig.~1 shows a 6-loop $\phi^4$
diagram whose counterterm cannot be obtained
by mere differentiation of Euler's formula,
$\ln\Gamma(1-z)=\gamma z+\sum_{n>1} \ze{n}z^n/n$,
from which all counterterms to 5 loops ultimately derive their
transcendentals.

The precise location of this obstacle has recently been confirmed by knot
theory~\cite{DK1,DK2,BKP}.  According to~\cite{DK1,DK2}
counterterms are rational if the link diagrams, encoding
the intertwining of momenta of the relevant Feynman diagrams, give
no non-trivial knots, when skeined. Conversely, knots obtained by skeining
link diagrams are in correspondence~\cite{BKP}
with distinct transcendental counterterms. The simplest example
is the $(2n-3,2)$ torus knot, with $2n-3$ crossings,
which is responsible for the appearance of $\ze{2n-3}$ in counter\-terms
at $n$ loops~\cite{DK2}.

Only special circumstances, such as a gauge symmetry, or supersymmetry,
lead one to expect lesser transcendental complexity.
For example, knot theory
relates~\cite{BDK} the cancellation of transcendentals in quenched QED
to the cancellation of subdivergences entailed by the Ward identity.
In $\phi^4$-theory~\cite{BKP}, where no such privileges apply,
one expects a new transcendental at the $n$-loop level
for {\em each\/} positive knot with $2n-3$ or $2n-4$ crossings, together
with products of lower-level transcendentals, corresponding to
factor knots~\cite{DK2}.
This is confirmed by the discovery~\cite{DK2} of the torus
knot $(4,3)=8_{19}$ in a positive 3-braid obtained from the 6-loop diagram
of Fig~1.
High-precision evaluation of this counterterm
reveals~\cite{BKP} it to be expressible in terms of
a double sum, previously encountered~\cite{1440,Z6}
in the $\ep$-expansion of two-loop diagrams
in $4-2\ep$ dimensions. At 7 loops one expects 4 new
transcendentals, corresponding to the 10-crossing knots
$10_{124}$, $10_{139}$, $10_{152}$, and a uniquely positive 11-crossing
hyperbolic knot, with braid word $\eltft$,
which was denoted by $11_{353}$ in~\cite{BKP}, on account of the
triple sum that it entails. The double sum associated with $(5,3)=10_{124}$
has also been identified~\cite{BKP}. Transcendentals associated with
knots with more crossings are given in~\cite{EUL,BK14,DK}.

The presence of further transcendentals, associated with the hyperbolic knots
$10_{139}$ and $10_{152}$, has been detected in only three highly complex
non-planar 7-loop diagrams~\cite{BKP}.
The precise multiple sums that are involved have not yet been identified,
since we have at present only 10-digit accuracy for the relevant
counterterms,
which is far less than that achieved for the conclusive identification
of the new transcendentals arising from $8_{19}$, $10_{124}$ and $11_{353}$.

In this paper we address the following questions:
\begin{enumerate}
\item[Q1] What is the simplest type of analytical structure
that produces non-zeta transcendentals in counterterms?
\item[Q2] Which anomalous dimensions entail such a structure?
\item[Q3] Are the resulting new
transcendentals at least as sparse as knot theory requires?
\item[Q4] How can one calculate the rational coefficients
that multiply the transcendentals in the anomalous dimensions?
\end{enumerate}

In brief, our answers are as follows:
\begin{enumerate}
\item[A1] Double sums, associated with the knots
$8_{19}$ and $10_{124}$, and the triple sum associated with $11_{353}$,
occur in the $\ep$-expansions of Saalsch\"utzian $\hyp$ series
generated by diagrams reducible to two-loop form.
\item[A2] Double sums contribute to anomalous dimensions,
in the large-$N$ limit~\cite{LNF}, at ${\rm O}(1/N^3)$;
triple sums do not appear
until one goes to higher order in $1/N$.
\item[A3] Up to level 11, the non-zeta transcendentals that are generated
by $\hyp$ series are {\em sparser\/} than required
by the enumeration of positive knots with up to 11 crossings. In particular,
more complex analytical structures are needed to generate the
transcendental knot-numbers corresponding to $10_{139}$ and $10_{152}$.
\item[A4] Combining knot, field, number and group theory,
we obtain the exact rational coefficients of all transcendentals
that occur in anomalous dimensions at ${\rm O}(1/N^3)$,
up to level 11, corresponding to 11 loops in 4-dimensional theories
and 12 loops in 2-dimensional theories.
We exemplify the result in the simplest case of a supersymmetric
theory~\cite{et3}, which does not entail level-mixing. Other cases
lead to lengthier expressions, with no further analytical complexity.
\end{enumerate}

The remainder of the paper demonstrates these answers, as follows.

In Section~2 we derive and solve a pair of recurrence relations
for the two-loop two-point diagram with 3 dressed lines, corresponding
to arbitrary exponents $\al{n}$ of
dressed propagators $1/(p_n^2)^{\al{n}}$ carrying
momenta $p_n$. The solution, in terms of a pair of $\hyp$ series,
is sufficient to determine anomalous dimensions to all orders
at ${\rm O}(1/N^3)$. {}From it, we extract a new class of zeta-reducible
two-point integrals.

Section~3 reviews previous work on $1/N$ expansions of critical exponents
and establishes the connection between ${\rm O}(1/N^3)$ results and a specific
case of the hypergeometric solution of Section~2.

Section~4 uses the wreath product group, $S_3\wr Z_2$,
of transformations of Saalsch\"utzian $\hyp$ series~\cite{DJB},
which enables us to enumerate
the minimal set of Taylor coefficients in the expansion of the two-loop
diagram that are not reducible to zetas. This enumeration is compared
with expectations based on the relation between counterterms and
positive knots~\cite{DK1,DK2,BKP}. Detailed analysis confirms these
expectations and provides further information about the relations between
knots and numbers, realized by field theory. We obtain
a convenient all-order reduction, to non-alternating double sums,
of the integral that is the source of all non-zeta terms in ${\rm O}(1/N^3)$
anomalous dimensions.

Section~5 exploits these results to obtain $\ep$-expansions of critical
exponents, and the related coupling-constant expansions of anomalous
dimensions. In the case of the 2-dimensional supersymmetric $\sigma$-model,
for which analytical expansions are the most compact,
we give the explicit reduction of a 12-loop result to
$\{\ze{n}\mid 3\leq n\leq11\}$, together with
the double sums $\ze{5,3}=\sum_{m>n>0}1/m^5n^3$ and
 $\ze{7,3}=\sum_{m>n>0}1/m^7n^3$, at levels 8 and 10.
Extending the analysis to 24 loops, we investigate the utility
of Pad\'e resummation in the $\sigma$ model, its supersymmetric extension,
the Gross-Neveu model, and $\phi^4$ theory.

Section~6 gives our conclusions.

\section{Beyond the triangle and uniqueness relations}

We begin by considering
the two-loop integral
\begin{equation}
I_6(\al1,\al2,\al3,\al4,\al5,\al6)
=\frac{p^{2(\mu-\al6)}}{\pi^{2\mu}}
\int\int\frac{\rd^{2\mu}k\,\rd^{2\mu}l}{
(k-p)^{2\al1}
(l-p)^{2\al2}
(k-l)^{2\al3}
l^{2\al4}
k^{2\al5}}\,,
\label{I6}
\end{equation}
with
$\al6\equiv3\mu-\al1-\al2-\al3-\al4-\al5$
determining the dependence
on the number, $d=2\mu$, of (euclidean) dimensions.
$I_6$ is clearly independent
of the norm of the external momentum $p$. Remarkably, it
has a 1,440-element symmetry group~\cite{1440}, corresponding
to all permutations of 6 linear combinations~\cite{Z6} of its arguments,
combined with the total reflection $\al{n}\to\ba{n}\equiv\mu-\al{n}$.
A function that is invariant under these $S_6\times Z_2$
transformations is
\begin{equation}
\overline{I}_6(\al1,\al2,\al3,\al4,\al5,\al6)
\equiv\frac{(2\mu-3)\left[G(1)\right]^2}
{\prod_{n=1}^{10}\left[G(\al{n})\right]^{1/2}}
\,I_6(\al1,\al2,\al3,\al4,\al5,\al6)\,,
\label{inv}
\end{equation}
where $G(\al{n})\equiv\Gamma(\ba{n})/\Gamma(\al{n})$,
and the auxiliary variables,
$\al7\equiv\al1+\al5-\ba6$,
$\al8\equiv\al2+\al4-\ba6$,
$\al9\equiv\al4+\al5-\ba3$,
$\al{10}\equiv\al1+\al2-\ba3$,
are associated with the vertices
of the tetrahedral vacuum diagram~\cite{1440}
of Fig.~2b, formed by joining the external lines of Fig.~2a.

Practically everything that is known about
Laurent expansions of multi-loop diagrams
can be derived from expanding\Eq{I6}, in $\ep=2-\mu$,
with arguments differing from integers by multiples of
$\ep$ resulting from integrations by parts that collapse
more complex diagrams to combinations of
diagrams of two-loop form, with self-energy
insertions. For example, there is only one~\cite{CT} independent
3-loop diagram
that cannot be reduced to combinations of\Eq{I6},
and it enters 4-loop
counterterms only at ${\rm O}(1/\ep)$, where its value, $20\ze5$, is the same
as for other 3-loop diagrams, which can be so reduced.
Moreover the expansion of\Eq{I6} is easily achieved to the level
of $\ze5$, which is the highest-level transcendental that enters
4-loop counterterms, corresponding to the torus knot $(5,2)$.

Group theory~\cite{1440,Z6} greatly reduces the burden of expanding $I_6$.
Only at the level of $\ze{13}$ does one encounter
an expansion coefficient of the six-argument function\Eq{inv} that
cannot be determined by expanding the four-argument function
\begin{equation}
I_4(\alpha,\beta,\gamma,\delta)\equiv I_6(\alpha,\beta,\gamma,1,1,
2\mu-\delta-2)\,;\quad\mu=\alpha+\beta+\gamma-\delta\,,
\label{I4}
\end{equation}
whose expansion is therefore sufficient to identify
transcendentals in $I_6$
corresponding to knots with up to 12 crossings.
Triangle relations~\cite{CT}, or equivalently uniqueness
relations~\cite{UNI}, reduce $I_4$ to $\Gamma$ functions
when any element of $\{\gamma,\,\delta,\,2\mu-\gamma-2,\,
2\mu-\delta-2\}$ is equal to unity, but such zeta-reducible
cases are insufficient to expand $I_6$ beyond the level of $\ze7$.
We shall expand it to level 11 and study the non-zeta transcendentals,
corresponding to positive knots with up to 11 crossings, sufficient
for 11 loops at ${\rm O}(1/N^3)$ in 4-dimensional field theory, and 12 loops
in 2-dimensional theories.

We proceed as follows. First we solve a master recurrence relation,
obtaining $I_4$ as a pair of $\hyp$ series.
This provides a new, two-argument, zeta-reducible result.
Reduction of the large-$N$ results of Section~3 to hypergeometric
series enables us to exploit the wreath-product group $S_3\wr Z_2$~\cite{DJB}
to study the knot theory of Section~4. In Section~5 we compute anomalous
dimensions to 12 loops, analytically, and to 24 loops, numerically.

\subsection{Solution of the master recurrence relation}

Operating on the integrand of\Eq{I6} with $k_\mu(\partial/\partial k_\mu)$
we obtain, from integration by parts,
\begin{eqnarray}
&&\alpha\,I_4(\alpha+1,\beta,\gamma,\delta+1)-
(\alpha+\gamma+2-2\mu)I_4(\alpha,\beta,\gamma,\delta)\nonumber\\
&&\quad{}=\gamma\,G(1,\delta+1)\left(\frac{\alpha\,G(\alpha+1,\gamma+1)}
{\delta-\beta+1}-G(\beta,\gamma+1)\right)\,,
\label{alp}
\end{eqnarray}
where $G(\al1,\al2)\equiv G(\al1)G(\al2)G(\ba1+\ba2)$
is the general one-loop integral, obtained by multiplication of
propagators in $x$-space.
The recurrence relation in $\beta$ is obtained
from\Eq{alp} by $\alpha\leftrightarrow\beta$ symmetry.

We now operate on integral\Eq{I6} with $(\partial/\partial p_\mu)^2$
and use\Eq{alp} to restore $\alpha$
and $\beta$ to their original values, to obtain a recurrence relation
in $\gamma$:
\begin{eqnarray}
&&(\gamma+2-2\mu)I_4(\alpha,\beta,\gamma,\delta)
+\frac{(\alpha+\gamma+1-2\mu)(\beta+\gamma+1-2\mu)}
{(\gamma+1-\mu)}
I_4(\alpha,\beta,\gamma-1,\delta-1)
\nonumber\\
&&\quad{}=\gamma(\delta+\gamma+1-2\mu)G(1,\delta)
\left(\frac{G(\alpha,\gamma+1)}{\delta-\beta}
+\frac{G(\beta,\gamma+1)}{\delta-\alpha}\right)\,.
\label{gap}
\end{eqnarray}

{}From\Eqq{alp}{gap} {\em all\/} other
identities can be obtained systematically. One can now forget
about triangle or uniqueness relations; they contain no further
information about $I_4$.
The recurrence relations are solved by finding a function, $S$,
with the properties:
\begin{eqnarray}
S(a,b,c,d)&=&S(b,a,c,d)\,,\label{sab}\\
S(c,d,a,b)+S(a,b,c,d)&=&0\,,\label{scd}\\
a(a+b+c+d)S(a,b,c,d)&=&a+b+c+d+(a+c)(a+d)S(a-1,b,c,d)\,.\label{src}
\end{eqnarray}
Given such a function, it is straightforward to verify that\Eqq{alp}{gap}
are satisfied by the Ansatz
\begin{equation}
\frac{I_4(\alpha,\beta,\gamma,\delta)}
{\gamma\,\delta\, G(1,\delta+1)}
=\frac{G(\alpha,\gamma+1)}{2\mu-3}
S(\mu-\alpha-1,\beta-1,\mu+\alpha-\delta-2,\delta-\beta)
+(\alpha\leftrightarrow\beta)\,,
\label{I4is}
\end{equation}
where $2\mu=2(\alpha+\beta+\gamma-\delta)$ is the number of dimensions.

Now the problem
is reduced to solving the master recurrence relation\Eq{src},
subject to symmetry properties\Eqq{sab}{scd}, which generate
all other recurrence relations. The solution involves
a $\hyp$
hypergeometric series\footnote{Reduction of a non-trivial two-loop
diagram to $\hyp$ series was first achieved in~\cite{SJH}.},
with one restriction on the 5 parameters,
which permits transformations discovered by
Saalsch\"utz~\cite{SAA,GHH}.
Scouring standard texts~\cite{WNB,LJS} and
formularies~\cite{ERH,BAT,PBM} for relevant properties of
Saalsch\"utzian $\hyp$ series, we find that
they are all contained in Hardy's elegant
study~\cite{GHH} of chapter XII of Ramanujan's notebook.
Defining the series~\cite{DJB}
\begin{equation}
F(a,b,c,d)\equiv
\sum_{n=1}^\infty\frac{(-a)_n(-b)_n}
{(1+c)_n(1+d)_n}
=\hyp\left[\begin{array}{l}-a,-b,1\\1+c,1+d\end{array}
;1\right]-1\,,
\label{fis}
\end{equation}
we encapsulate its properties as follows:
\begin{eqnarray}
F(a,b,c,d)&=&F(b,a,c,d)\,,\label{fab}\\
(a+c)(b+c)F(a,b,c,d)&=&a\,b\,F(-b-c,-a-c,c,a+b+c+d)\,,\label{fsa}\,\\
(a+c)(a+d)F(a,b,c,d)&=&a b +a(a+b+c+d)F(a-1,b,c,d)\,,\label{frc}\\
F(c,d,a,b)+F(a,b,c,d)&=&H(a,b,c,d)-1,\label{fcd}\\
F(a-b,-2b,a+b,2b)&=&2b\left[\psi(1+a)-\psi(1+2a)
-\psi(1+b)+\psi(1+2b)\right]\,,
\label{hdy}
\end{eqnarray}
where $\psi(z)\equiv\Gamma^\prime(z)/\Gamma(z)$, and
\begin{equation}
H(a,b,c,d)\equiv
\frac{
\Gamma(1+a)
\Gamma(1+b)
\Gamma(1+c)
\Gamma(1+d)
\Gamma(1+a+b+c+d)}{
\Gamma(1+a+c)
\Gamma(1+a+d)
\Gamma(1+b+c)
\Gamma(1+b+d)}
\label{hdef}
\end{equation}
generates all known special cases, save that of\Eq{hdy}.

A solution to~(\ref{sab},\ref{scd},\ref{src}),
and hence to\Eqq{alp}{gap}, is given by
\begin{equation}
S(a,b,c,d)=\frac{\pi\cot\pi c}{H(a,b,c,d)}
-\frac{1}{c}-\frac{b+c}{b c}F(a+c,-b,-c,b+d)\,,
\label{sis}
\end{equation}
where\Eq{fsa} gives\Eq{sab},\Eq{frc} gives\Eq{src}, and\Eq{fcd}
gives\Eq{scd}, via the elementary identity
\begin{equation}
\cot\pi b+\cot\pi c=\left[\frac{1}{\pi b}
+\frac{1}{\pi c}\right]H(a,b,c,d)H(a+c,-b,-c,b+d)\,.
\label{ident}
\end{equation}

\subsection{A new case of zeta-reducibility}

Properties\Eqq{fsa}{fcd} ensure that $F(a,b,c,d)$ is reducible to $\Gamma$
functions, or zero, when any element of $\{a,b,c,d,a+b+c+d\}$ vanishes.
Hence\Eq{sis} makes $S(a,b,c,d)$ reducible when any element of
$\{a+c,a+d,b+c,b+d\}$
vanishes. Hence\Eq{I4is} makes $I_4(\alpha,\beta,\gamma,\delta)$
reducible when any element of $\{\gamma,\,\delta,\,2\mu-\gamma-2,\,
2\mu-\delta-2\}$ is equal to unity. Evaluating these special instances,
we obtain agreement with known results from
the triangle and uniqueness relations,
thereby checking that\Eqq{I4is}{sis}
give the correct solution to\Eqq{alp}{gap}, i.e.\ that no homogeneous
solution to the recurrence relations has been omitted. We have also
checked\Eqq{I4is}{sis} in instances that are not reducible,
by comparing results from single sums of the form\Eq{fis}
with the far less convenient
double\footnote{A transformation from double sums to single sums
was found in~\cite{AVK}, following communication that we had obtained
$\hyp$ series by solving recurrence relations.}
sums that result from Gegenbauer
polynomial~\cite{GPX} methods.
Likewise we find numerical
agreement with a more cumbersome reduction to ${}_4F_3$ series
given in~\cite{Z6}.

There is one more case in which $I_4$ has an expansion
that is reducible to zetas. It results from\Eq{hdy},
obtained from identity~8.4 of~\cite{GHH}, which
gives the 4-dimensional result
\begin{equation}
\overline{I}_6(1+u,1+u,1-u-v,1,1,1-u+v)=
2\sum_{s=1}^\infty\ze{2s+1}(1-4^{-s})
\frac{(u+v)^{2s}-(u-v)^{2s}}{u v}\,.
\label{new}
\end{equation}
Single-argument special cases of this proved useful
in~\cite{KAZ,1440,Z6,LLP}.
The two-argument result is new and enables us
to derive, by simple algebra,
the $S_6\times Z_2$-invariant expansion of\Eq{inv} that was
given to level 9 in~\cite{Z6}, where only one non-zeta transcendental was
encountered. This is shown in~\cite{DK2,BKP} to be
associated with the 8-crossing knot $(4,3)=8_{19}$.
At level 9 we achieve reduction to zetas without the need of the numerical
searching that was used in~\cite{Z6}.
What happens at higher levels is the subject of Section~4.

\section{Critical exponents at large $N$}

Before searching for level-$10$ knot-transcendentals in the expansion
of\Eq{I4is}, we review the manner in which such a two-loop integral
enters the computation of critical exponents in the large-$N$ limit.

Results from multi-loop perturbation theory are important for
high-precision computation
of critical exponents in statistical physics,
allowing comparison with accurate experimental values.
A widely studied example is the Heisenberg ferromagnet.
It undergoes a phase transition, from a disordered to an ordered phase,
with properties of the transition characterized solely by critical exponents.
The field theories underlying the transition are the ${\rm O}(3)$
nonlinear $\sigma$-model, or $(\phi^2)^2$-theory with an ${\rm O}(3)$
symmetry, in $d=2\mu=3$ dimensions.
The $\beta$-function of each theory has a non-trivial zero,
which is identified as the phase transition of the Heisenberg model.
At their fixed points, the two theories are equivalent,
or said to be in the same universality class.
In other words, an exponent derived from one theory has the
same value as that derived from the other.
Consequently, analytic calculations may be performed in either model to
predict values for the exponents.

In computing the exponents from $(\phi^2)^2$-theory,
a model with a more general ${\rm O}(N)$ symmetry may be considered,
either directly in three dimensions~\cite{Bak},
or near four dimensions~\cite{PHI,Bre,Vla}.
We describe the procedure in the latter case, as it entails
the expansion in $\ep$ of diagrams calculated in $d=2\mu=4-2\ep$
dimensions, where non-zeta terms will occur, via the expansion of\Eq{I4is}.
First, the theory is renormalized, to some order in perturbation theory,
at $d=4$ dimensions, using dimensional regularization and minimal subtraction
of powers of $1/\ep=2/(4-d)$.
{}From the $\beta$-function, one then deduces
the location of the non-trivial $d$-dimensional fixed point, $g_c(\ep)$,
as a power series in $\ep$. Knowledge of the wavefunction and mass anomalous
dimensions, $\gamma_2(g)$ and $\gamma_m(g)$,
to corresponding orders in the coupling constant $g$, then yields
the critical exponents~\cite{PHI} $\eta=2\gamma_2(g_c)$,
$1/\nu\equiv2\lambda=2[1-\gamma_m(g_c)]$,
and $\omega=2\beta^\prime(g_c)$
as power series in $\ep$.
Values for three dimensions are obtained by improving the
convergence of these series, using methods of accelerated convergence,
after which one sets $\ep=\frac12$.
This procedure was applied initially at the three-loop level,
yielding values in impressive agreement with experiment~\cite{LeG}.
Later, four- and five-loop results were used to improve
the accuracy~\cite{Vla,Gor}.
In addition to the $N=3$ case of the Heisenberg ferromagnet,
${\rm O}(N)$-symmetric $(\phi^2)^2$-theory
underlies the critical behaviour of polymers, the Ising model,
and superfluid Bose liquids, at $N=0$, $1$
and $2$, respectively. Moreover,
the $N=1$ case of the three-dimensional Ising model
is in the same universality class as the deconfinement~\cite{YM}
transition of pure ${\rm SU}(2)$ Yang-Mills theory in
$d=4$ spacetime dimensions.

As an alternative to the $\ep$-expansion at fixed $N$,
one can develop a large-$N$ expansion at fixed $\ep$.
In this approach, the critical exponents
are expanded in powers of $1/N$, with the coefficients in
$\eta=\sum_k\eta_k/N^k$ determined for a particular
value of $\ep$, or for an arbitrary value.
With a reasonable number of terms, one hopes to
improve the convergence of the series, in a manner similar to that for
the $\ep$-expansion, before setting $N=3$, or some other low value.
Initially, this approach was developed for the ${\rm O}(N)$ $\sigma$-model
in strictly three dimensions, with
exponents determined to ${\rm O}(1/N^2)$~\cite{Okabe,Kon}.
Later these results were extended in an impressive series of
papers~\cite{Vas1,Vas2,Vas3} by Vasil'ev and co-workers,
who obtained $\eta$ at ${\rm O}(1/N^3)$ and $1/\nu$ at ${\rm O}(1/N^2)$,
in an arbitrary spacetime dimension.
The method is based on ideas from~\cite{Bray},
and on the application of the conformal bootstrap programme
of~\cite{Mig,Pol,Par}.
It is possible to compute in $d=2\mu$ dimensions
by exploiting the conformal symmetry
that exists at the fixed point, making use of so-called uniqueness
relations for conformal integration,
to determine the values of three- and four-loop diagrams that
occur in the corrections to the Dyson-Schwinger equations and hence to
the critical exponents.
An obvious check on these calculations is the reproduction
of previous results on restriction to three dimensions.

There are further checks on the $d$-dimensional results at large-$N$,
based on the equivalence of the two underlying field theories at their
fixed points. For concreteness, we recall that the lagrangians are
\begin{equation}
{\cal L}=\df{1}{2}(\partial_\mu\phi_i)^2+\df{1}{2}\sigma(\phi_i^2-1/g)
\,,\label{L1}
\end{equation}
for the $\sigma$-model, which is perturbatively renormalizable
only in two dimensions, and
\begin{equation}
{\cal L}=\df{1}{2}(\partial_\mu\phi_i)^2+\df{1}{2}\sigma(\phi_i^2-\sigma/h)
\,,\label{L2}
\end{equation}
for $(\phi^2)^2$-theory. The flavour index $i$ is summed from 1 to $N$,
and $\sigma$ acts as a Lagrange-multiplier field in\Eq{L1},
but as an auxiliary field in\Eq{L2}, with $g$ and $h$ as
the respective coupling constants.
The $(\phi^2)^2$ formulation\Eq{L2} was used in the initial study of the
relation of knots to counterterms~\cite{DK1,DK2,BKP}.
To obtain exponents from the $\sigma$-model formulation\Eq{L1},
one follows the procedure described above for $(\phi^2)^2$-theory
in $d$-dimensions, but sets $d=2-2\ep$ to obtain $\ep$-expansions
that yield three-dimensional values at $\ep=-\frac12$.
In the case of the $\sigma$-model, it is the slope of the
$\beta$-function that now determines the critical exponent $2\lambda=1/\nu$.

Highly non-trivial checks result from comparing large-$N$ expansions
of critical exponents, at arbitrary $\ep$, with $\ep$-expansions
at arbitrary $N$, obtained from either field theory.
Agreement was demonstrated in~\cite{Vas1,Vas2,Vas3} up to
orders then known in perturbation theory.
Now that corrected~\cite{PHI} five-loop results have recently been given for
$\phi^4$-theory, we have checked all available
large-$N$ results for the exponents, namely
$\eta$ to ${\rm O}(1/N^3)$~\cite{Vas3},
$2\lambda\equiv1/\nu$ to ${\rm O}(1/N^2)$~\cite{Vas2},
and $\omega$ to
${\rm O}(1/N^2)$, where we calculated $\omega$
by extending the techniques of~\cite{Vas2} to include
corrections to the asymptotic scaling forms of the propagators due to
insertion of an operator with dimension $(\mu-2)$,
obtaining a result that is reported
in Section~5.1, where
full agreement with the 5-loop results of~\cite{PHI} is found.

Given this impressive agreement between all-order results at large $N$
and 5-loop perturbation theory at fixed $N$, we may confidently
take the $\ep$-expansion of the ${\rm O}(1/N^3)$ result for $\eta$~\cite{Vas3}
as containing transcendentals that will be encountered in the loop
expansions of four-dimensional $\phi^4$-theory, and the
two-dimensional $\sigma$-model, beyond the levels
thus far computed in fixed-$N$ minimally-subtracted perturbation theory.
Hence large-$N$ results provide a window on transcendentals
from knots with many crossings.

It is at ${\rm O}(1/N^3)$ in the critical exponent $\eta$ that one first
encounters an integral, in $d=2\mu$ dimensions, which has resisted all
attempts at reduction to $\Gamma$ functions and their logarithmic
derivatives, the polygamma functions $\psi^{(n)}(z)\equiv(\rd/\rd
z)^{n+1}\ln\Gamma(z)$. {}From the point of view of the authors
of~\cite{Vas3}, this irreducibility was understandably annoying.
{}From our
point of view, it is a great bonus, since it leads us into the domain of
knots more complex than the
$(2n-3,2)$ torus knots that produce $\ze{2n-3}$ at $n$ loops in
$\phi^4$-theory~\cite{DK2,BKP}. The irreducible term encountered
at ${\rm O}(1/N^3)$ was denoted by $I(\mu)$ in~\cite{Vas3} and is
the logarithmic derivative of a two-loop self-energy diagram,
$\Pi(\mu,\Delta)$, with respect to the exponent, $\Delta$,
of the completely internal line. In terms
of the two-loop momentum-space integral\Eq{I4}, it is defined by
\begin{equation}
I(\mu)\equiv\left.\frac{\rd\ln\Pi(\mu,\Delta)}{\rd \Delta}\right|_{\Delta=0}
\,;\quad
\Pi(\mu,\Delta)\equiv I_4(\mu-1,\mu-1,\mu-1+\Delta,2\mu-3+\Delta)\,,
\label{Pi}
\end{equation}
and results from non-planar diagrams in the skeleton
Dyson-Schwinger equation for the $\sigma\phi^2$ vertex~\cite{Vas3}.
The same integral occurs, with different $\mu$-dependent coefficients,
in $\eta_3$ for the Gross-Neveu model~\cite{gn3,Der}
and for the supersymmetric $\sigma$-model~\cite{et3}.

There is a simple rule (which will be derived later) relating
the number of loops in perturbative expansions to $\ep$-expansions of
$I(\mu)$: for 2-dimensional theories, the term of ${\rm O}(\ep^n)$
in the expansion of $\ep I(1-\ep)$ is of level $n$ and first appears at
$n+1$ loops; for 4-dimensional theories, this level-$n$ term
first appears at $n$ loops.

\section{Non-zeta terms and positive knots}

Now we have shown how large-$N$ anomalous dimensions involve
the expansion of\Eq{I4is}, at ${\rm O}(1/N^3)$, we proceed to enumerate
the possible non-zeta terms in the expansion of\Eq{fis}
and compare the tally with the numbers allowed by knot theory
in the $\ep$-expansion of\Eq{Pi}.

\subsection{The wreath-product group $S_3\wr Z_2$}

It is clear that~(\ref{fab},\ref{fsa},\ref{fcd}) relate a large number
of $\hyp$ series, with transformed arguments.
In fact the group of transformations has 72 elements, as can be seen
by considering its operation on the matrix
\begin{equation}
M\equiv\left[\begin{array}{ccc}
-(b+d)&-(b+c)&a\\
-(a+d)&-(a+c)&b\\
c&d&(a+b+c+d)\end{array}\right]\,.
\label{mat}
\end{equation}
The symmetry\Eq{fab} corresponds to exchange of rows 1 and 2;
Saalsch\"utz's transformation\Eq{fsa} to exchange of columns 2 and 3.
Combining these with the transformation entailed in relation\Eq{fcd},
we may transpose the matrix, permute its rows, and
permute its columns. The group of such transformations is the wreath
product $S_3\wr Z_2$~\cite{DJB}.
To generate its polynomial invariants, we define
\begin{eqnarray}
&&\lambda_1=\df13(+a-2b-c-d)\,;
\quad\lambda_2=\df13(-2a+b-c-d)\,;\quad
\lambda_3=-\lambda_1-\lambda_2\,;\nonumber\\
&&\mu_1=\df13(-a-b+c-2d)\,;\quad \mu_2=\df13(-a-b-2c+d)\,;
\quad\mu_3=-\mu_1-\mu_2\,;\nonumber\\
&&N_n^\pm=\sum_{i=1}^3 \frac{\mu_i^n\pm\lambda_i^n}{2}\,;\quad
N(p_2,m_2,p_3,m_3)=
\left[N_2^+\right]^{p_2}
\left[N_2^-\right]^{m_2}
\left[N_3^+\right]^{p_3}
\left[N_3^-\right]^{m_3}\,.
\label{pm}
\end{eqnarray}
By construction, the matrix elements of\Eq{mat} are
$M_{i,j}=\lambda_i+\mu_j$, and $N_1^\pm$ vanish.
With arguments of order $\ep$, a complete set of linearly independent
wreath-product invariants, at ${\rm O}(\ep^n)$, is given by
\begin{equation}
\left\{\,N(p_2,m_2,p_3,m_3)\mid n=2(p_2+m_2)+3(p_3+m_3)\,,
(-1)^{m_2}=(-1)^{m_3}\,\right\}\,.
\label{comp}
\end{equation}

It remains to relate the $\hyp$ series\Eq{fis}
to an invariant function, whose expansion
in terms of\Eq{comp} contains all the non-zeta terms.
To achieve this, we define
\begin{eqnarray}
\overline{F}(a,b,c,d)&=&
\frac{(ab-cd)F(a,b,c,d)+ab}{abcd(a+b+c+d)}
-\frac{\Gamma(c)\Gamma(d)\Gamma(a+b+c+d)}{\Gamma(1+a+c+d)\Gamma(1+b+c+d)}\,,
\label{fb}\\
\disp{a}{b}{c}{d}&=&
\overline{F}(c,d,a,b)-\overline{F}(a,b,c,d)\,,
\label{ab}\\
W(a,b,c,d)&=&
\overline{F}(a,b,c,d)
+\frac16\disp{-a-c}{-a-d}{a}{a+b+c+d}
+\frac16\disp{-b-c}{-b-d}{b}{a+b+c+d}\nonumber\\
&+&\frac12\disp{a}{b}{c}{d}
+\frac16\disp{-a-c}{-b-c}{c}{a+b+c+d}
+\frac16\disp{-a-d}{-b-d}{d}{a+b+c+d}.\label{wp}
\end{eqnarray}
By construction,\Eq{fb} is regular when any element of
$\{a,b,c,d,a+b+c+d\}$ vanishes,\Eq{fcd} reduces\Eq{ab} to $\Gamma$
functions, and\Eq{wp} is a wreath-product invariant, which is
reducible to $\Gamma$ functions if $N_2^-=0$ or $N_3^+=N_3^-=0$.
The vanishing of $N_2^-=ab-cd$ clearly removes $F$ from\Eq{fb}, while
the vanishing of both $N_3^+$ and $N_3^-$ corresponds to the two-argument
zeta-reducible case in\Eq{hdy}. To remove from $W$ the trivial
case $ab=cd$, we define
\begin{eqnarray}
\overline{W}(a,b,c,d)&=&W(a,b,c,d)
-\sum_{n=1}^3 \frac{
 T(M_{1,n},M_{2,n},M_{3,n})
+T(M_{n,1},M_{n,2},M_{n,3})}{6}\,,\label{sub}\\
T(x,y,z)&=&\frac{1}{x y z}-\frac{\Gamma(x)\Gamma(y)\Gamma(z)}
{\Gamma_+(x,y,z)\Gamma_-(x,y,z)}\,,\label{tri}\\
\Gamma_\pm(x,y,z)&=&\Gamma\left(1+\df12(x+y+z)
\pm\df12\left[x^2+y^2+z^2-2x y-2y z-2z x\right]^{1/2}\right)\,,
\label{root}
\end{eqnarray}
with a subtraction in\Eq{sub} that is also a wreath-product invariant,
thanks to the symmetry of\Eq{tri}. When $N_2^-=0$, all instances
of the square root in\Eq{root} give rational differences of matrix elements,
making $\overline{W}$ vanish. In any case, expanding\Eq{tri} removes
the square root, producing the invariants\Eq{comp} at ${\rm O}(\ep^n)$.

The expansion of $\overline{W}$,
in terms of the invariants\Eq{comp},
contains only terms with $m_2>0$ and even values of $m_2+m_3$.
Moreover, those with $p_3=m_3=0$ are zeta-reducible.
The leading term occurs at level 7 and is determined
by Hardy's result\Eq{hdy}, which gives
\begin{equation}
\overline{W}(a,b,c,d)=
\left[
-\df{75}{16}\ze7
          +3\ze5\ze2
  -\df{3}{2}\ze4\ze3
\right](ab-cd)^2+{\rm O}(\ep^5)\,,
\label{lead}
\end{equation}
with transcendentals of level $n+3$ entering at ${\rm O}(\ep^n)$.

It is now straightforward to count the expansion coefficients,
at any given level, and to determine how many are zeta-reducible.
Up to level 20, the tally is
\begin{equation}
\begin{array}{lcccccccccccccc}
\mbox{level:}&
\,7\,&\,8\,&\,9\,&10\,&11\,&12\,&13\,&14\,&15\,&16\,&17\,&18\,&19\,&20\\
\mbox{coefficients:}&
                 1 & 1 & 1 & 2 & 3 & 3 & 5 & 6 & 7 & 9 & 11 & 12 & 16 & 18\\
\mbox{zeta-reducible:}\quad&
                 1 & 0 & 1 & 0 & 2 & 0 & 2 & 0 & 3 & 0 &  3 &  0 &  4 &  0\\
\mbox{non-zeta:}&
                 0 & 1 & 0 & 2 & 1 & 3 & 3 & 6 & 4 & 9 &  8 & 12 & 12 & 18
\end{array}\label{non}
\end{equation}
whose non-zeta terms we expect to involve transcendentals associated
with knots more complex than the $(2n-3,2)$ knot that generates
$\ze{2n-3}$ in subdivergence-free counterterms of 4-dimensional theories
at $n$ loops.

\subsection{Expectations from knot theory}

To see what knot theory has to tell us in this context, we consider
dressings of the three-loop tetrahedron vacuum diagram by chains
of one-loop self-energy insertions, since counterterms with this structure
may be obtained from the $\ep$-expansion of the master two-loop diagram,
by the method of infrared rearrangement~\cite{IRA}.

Unadorned, the tetrahedron delivers $\ze3$,
which corresponds to the trefoil knot
\cite{DK1,DK2}. For example, the 3-loop coefficient
of the $\beta$-function of $\phi^4$ theory,
in 4 dimensions, receives a scheme-independent contribution $6\ze3$
from the counterterm that cancels the overall divergence of the
subdivergence-free four-point function obtained by attaching
external lines to the vertices of the tetrahedron of Fig.~2b.
Its evaluation is straightforward, since it is merely the finite value
of the diagram of Fig.~2a, with massless propagators, $1/p_n^2$,
and unit external momentum, which is easily obtained by
the triangle rule~\cite{CT},
or the method of uniqueness~\cite{UNI}.

Now we dress this three-loop vacuum diagram
with chains of one-loop self-energy insertions,
thus extending the domain of investigation to counterterm diagrams
with subdivergences. Since we are interested, at this stage,
in the momentum flow, we do not specify the particle content
of a specific field theory, though we have in mind our experience~\cite{BKP}
of $(\phi^2)^2$ theory, with an ${\rm O}(N)$ symmetry group,
whose self-energy insertions would involve
shrinking some of the lines in Figs.~3--8, without affecting
the flow patterns.

Such self-energy insertions modify the powers of propagators
in the 3-loop vacuum diagram, so that we are now dealing with transcendentals
generated by closing a two-loop two-point diagram that contains
non-integer lines, which\Eq{Pi} confirms as the arena explored by
large-$N$ analyses.

Experience with diagrams with subdivergences is reported
in~\cite{DK1}, where it was observed that the dressing of ladder
topologies turns rational~\cite{DKT} counterterms into counterterms containing
both odd and even zetas. The occurrence of odd zetas
in subdivergence-free diagrams was studied
in~\cite{DK1,DK2,BKP}, where
$\ze{2n+1}$ was associated with the torus knot $(2n+1,2)$.
The even zetas, $\{\ze{2n}\mid n\geq2\}$, are restricted to diagrams with
subdivergences, leading to
link diagrams with non-zero writhe numbers~\cite{DK1,DK2}.
In general, a dressed ladder diagram with $n\geq4$ loops may
contribute $\{\ze{k}\mid 2<k<n\}$
to the overall counterterm, after subtraction of subdivergences
from dressed ladders in 4 dimensions.
For the dressed tetrahedron, we find that
non-zeta transcendentals of level $n\geq8$ occur at $n$ loops
for 4-dimensional field theories, and $n+1$ loops for 2-dimensional theories.

The question now at issue is this: how does the insertion of
self-energy dressings in the 3-loop tetrahedron turn a zeta-reducible
counterterm into one that is not reducible to the torus knots
$(2n+1,2)$?
We shall exhibit, in Figs.~3--8,
self-energy insertions that generate
the knots $8_{19}$, $10_{124}$, associated with irreducible~\cite{BBG}
double Euler sums~\cite{LE}, the knot $11_{353}$, associated with an
irreducible triple Euler sum~\cite{BG},
and a {\em pair\/} of 12-crossing knots,
whose appearance is associated with {\em alternating\/} double sums.

We start our investigation with Fig.~3a, where
three propagators are dressed. With
such a dressing we find
that all momentum routings yield link diagrams whose
skeining produces the double-sum knot $8_{19}$, familiar from the
6-loop subdivergence-free contributions to the $\beta$-function of
$\phi^4$ theory~\cite{BKP}.
Note also that this pattern
of insertions leaves no vertex, nor any triangle, unmodified,
and hence reduction to zetas by the method of uniqueness is no longer
possible.

Next, in Fig.~3b, we assign a 8-component link diagram to the 8-loop
Feynman diagram of Fig.~3a, following the methodology of~\cite{DK1,DK2}.
The momentum flows of
the Feynman diagram become components of the link diagram,
and each vertex in the diagram corresponds to a crossing
in the link diagram (though the converse is not always the case).
In Fig.~3b we see three rings generated by the three dressings
in the upper half of Fig.~3a. Their presence forces non-trivial
entanglement elsewhere in the link diagram. Having served that purpose
they play no further role, since the skeining process, associated with
removal of subdivergences, removes all trace of them from the final
knots that are in correspondence with the transcendentals
in the overall counterterm.
The entanglement generated by the remaining pair of insertions, in
the bottom right of Fig.~3a, is of the essence in this example.
These insertions sit on a propagator that necessarily carries the flow of
{\em two\/} of the three loop momenta of the tetrahedral skeleton, when each
of the upper propagators carries only a single loop momentum of the
tetrahedron.
To generate $8_{19}$, we have chosen one of the
dressings to have a non-local momentum flow, and one to have
a local flow. The local momentum flow
runs through only two propagators
(i.e.\ it remains in the smallest forest of the associated subdivergence,
in the language of renormalization theory).
If we had chosen a local flow for both self-energy insertions,
we would have generated a 2-braid knot, giving merely the zeta content
of the overall counterterm, whereas its non-zeta content corresponds to the
3-braid that is obtained by skeining Fig.~3b.

Next, we use skeining to reduce the 8-component link diagram of Fig.~3b
to the 3-component link diagram of Fig.~3c.
Three components are trivially removed
by skeining the upper rings of Fig.~3b to give curls that
can be untwisted; a further skeining and untwisting removes
all trace of the central propagator; a final skeining, at the bottom right,
generates a non-trivial entanglement. In Fig.~3c, a central dot
is marked, as an origin for reading the positive
braid word for the 10-crossing 3-component link,
$\sigma_1^3\sigma_2^2\sigma_1^{}\sigma_2^4$,
starting at bottom of the figure and reading anti-clockwise,
with a crossing of the outermost line over the next-to-outermost
denote by $\sigma_1$, and a crossing of the next-to-outermost over
the innermost by $\sigma_2$.

Finally, we perform two more skeinings to generate
the knot $8_{19}$ of Fig.~3d, with braid word
$\sigma_1^2\sigma_2^2\sigma_1^{}\sigma_2^3$.
Using the relation
$\sigma_1\sigma_2\sigma_1
=\sigma_2\sigma_1\sigma_2$, between braid group generators,
one may prove that
$\sigma_1^{2k}\sigma_2^2\sigma_1^{}\sigma_2^{2l+1}=
\sigma_1^{}\sigma_2^{2k+1}\sigma_1^{}\sigma_2^{2l+1}$
and hence obtain the more familiar braid words
$(\sigma_1^{}\sigma_2^3)^2=(\sigma_1^{}\sigma_2)^4$,
which show that $8_{19}$ is indeed the $(4,3)$ torus knot~\cite{VJ}.

To generate the next member of the series of non-zeta knots,
we proceed as illustrated in Fig.~4, repeating the methodology of Fig.~3.
Note that we add {\em two} more self-energy insertions
to Fig.~3a to obtain Fig.~4a: one spectates; the other participates
in the entanglement. Each is involved, since the spectator forces
a pair of tetrahedron loop momenta to flow through a chain of three
self-energy insertions. The 10-loop Feynman diagram of Fig.~4a then gives the
10-component link of Fig.~4b, which in turn skeins to
the 12-crossing 3-component link of Fig.~4c, with braid word
$\sigma_1^5\sigma_2^2\sigma_1^{}\sigma_2^4$. Two final skeinings (omitted
from Fig.~4) then deliver the knot
$\sigma_1^4\sigma_2^2\sigma_1^{}\sigma_2^3=
\sigma_1^{}\sigma_2^5\sigma_1^{}\sigma_2^3=10_{124}$, which is,
coincidentally, the $(5,3)$ torus knot.
Note that we again chose one dressing to have non-local
momentum flow. In the case of Fig.~4b it is the rightmost
self-energy insertion. We have verified that each of the other choices
leads to the same knot.
Moreover, choosing more dressings to have non-local flows does not
produce any more complicated knot. In particular the positive
hyperbolic knots $10_{139}$ and $10_{152}$ cannot be obtained
by such a process.

At 12 loops a new feature emerges: a {\em pair\/} of 12-crossing
knots is generated by self-energy insertions, as illustrated by
Figs.~5 and~6, which refer to the same Feynman diagram.
The momentum flow encoded by the link diagram of Fig.~5b gives
the 3-component link of Fig.~5c, which delivers the knot
$\sigma_1^6\sigma_2^2\sigma_1^{}\sigma_2^3=
\sigma_1^{}\sigma_2^7\sigma_1^{}\sigma_2^3$.
On the other hand, the momentum flow of Fig.~6b delivers a different knot:
$\sigma_1^4\sigma_2^2\sigma_1^{}\sigma_2^5=
\sigma_1^{}\sigma_2^5\sigma_1^{}\sigma_2^5$.
Other choices of routing and skeining
deliver no further non-zeta knots.

{}From explicit calculation of HOMFLY polynomials~\cite{VJ} of
positive braid words of length 12 we find that there are
precisely 7 positive non-torus knots with crossing number 12.
All are 3-braids and are listed in Table~1.
However, only the first two are obtained by dressing three propagators
of the tetrahedron.

{\bf Table~1}:
Positive prime knots related to multiple sums,
via field-theory counterterms.
\[\begin{array}{|l|l|l|}\hline
\mbox{crossings}&\mbox{knots}&\mbox{numbers}\\[3pt]\hline
2a+1
&\sigma_1^{2a+1}&\ze{2a+1}\\\hline
8
&\fk{}3{}3=8_{19}&N_{5,3}\\\hline
9
&\mbox{none}&\mbox{none}\\\hline
10
&\fk{}5{}3=10_{124}&N_{7,3}\\
&\fk{}333=10_{139}&?\\
&\fk2233=10_{152}&?\\\hline
11
&\eltft&N_{3,5,3}\\\hline
12
&\fk{}7{}3&N_{9,3}\\
&\fk{}5{}5&N_{7,5}-\frac{\pi^{12}}{2^5 10!}\\
&\fk{}353&?\\
&\fk{}335&?\\
&\fk2235&?\\
&\fk2334&?\\
&\fk3333&?\\\hline
\end{array}\]

Hence we expect that 12-loop counterterms, generated in the manner
of Figs.~5 and~6, will contain up to two double-sum level-12 transcendentals
that are not reducible to zetas. More generally, we expect
$[m/2]$ irreducibles of level $2m+4$, associated with
the knots
\begin{equation}
\{\sigma_1^{}\sigma_2^{2k+1}\sigma_1^{}\sigma_2^{2l+1}\mid
l+k=m\,,\,k\geq l\geq1\}\,.\label{knots}
\end{equation}
At levels 8 and 10, we find that
subdivergence-free diagrams deliver the knot numbers
$N_{5,3}$ and $N_{7,3}$, where~\cite{EUL}
\begin{equation}
N_{a,b}\equiv\sum_{m>n>0}\left(\frac{(-1)^m}{m^a n^b}-\frac{(-1)^n}{n^a m^b}
\right)\,.\label{nab}
\end{equation}
In~\cite{BKP} it was found that the knots $8_{19}$ and $10_{124}$
are associated with the combinations $29\ze8-12\ze{5,3}$
and $94\ze{10}-793\ze{7,3}$, in subdivergence-free counterterms,
where
\begin{equation}
\ze{a,b}\equiv\sum_{m>n>0}\frac{1}{m^a n^b}\label{zab}
\end{equation}
are non-alternating double sums. Subsequently it was realized that
the $\pi^2$ terms in these combinations are precisely those
generated by the alternating sums $N_{5,3}$ and $N_{7,3}$. In higher-loop
subdivergence-free diagrams we encounter $N_{9,3}$ and
$N_{7,5}-\frac{\pi^{12}}{2^5 10!}$.

In general, we conjecture that the
knots\Eq{knots} are associated, via subdivergence-free counterterms,
with the numbers
\begin{equation}
\{N_{2k+3,2l+1}\mid
l+k=m\,,\,k\geq l\geq1\}\label{numbers}
\end{equation}
modulo a rational multiple of $\pi^{2k+2l+4}$ that vanishes when $k=l$.

Analytical and numerical investigations~\cite{EUL} strongly
support the irreducibility of\Eq{numbers}. Moreover, it was
the methodology of Figs.~3-6 that led to the discovery that
double-sum irreducibles increase by the `rule of two' entailed by the
knots\Eq{knots}, with $[m/2]$ knot-numbers at level $2m+4$. This
is in marked contrast to the `rule of three' that governs the
non-alternating sums\Eq{zab}, of which $[(m+1)/3]$ are irreducible~\cite{BBG}
at level $2m+4$. {}From this we conclude that one should
look beyond non-alternating sums, when studying the transcendental
content of counterterms. Interestingly, it appears that non-alternating
sums do not inhabit a cosy world of their own: in the course of this
work one of us discovered~\cite{EUL} a four-fold non-alternating sum,
$\ze{4,4,2,2}=\sum_{k>l>m>n>0}1/k^4l^4m^2n^2$,
which is reducible to alternating double sums,
but not to non-alternating
double sums. This led to an enumeration~\cite{EUL} of irreducible
multiple sums that is much simpler than might have been suspected
when attention was restricted to non-alternating sums.

Having associated the knots\Eq{knots} with the numbers\Eq{numbers},
by consideration of the tetrahedron with three dressed
propagators, we now proceed to consider cases in which more than
three propagators are dressed, having in mind that we now probe
higher terms in the large-$N$ expansion than are entailed by the
${\rm O}(1/N^3)$ analysis that produces\Eq{Pi}. We expect to find the 4-braid
knot $\eltft\equiv11_{353}$,
associated~\cite{BKP} with a single~\cite{BG} irreducible
level-11 triple sum, which may be taken as $\ze{3,5,3}=
\sum_{l>m>n}1/l^3m^5n^3$.
It is also interesting to see whether the 3-braids
$10_{139}$ and $10_{152}$ emerge from complicated dressings of the tetrahedron.

Detailed investigation suggests that four dressed propagators
are insufficient to
generate $11_{353}$.
Investigating the five dressings of Fig.~7a, we obtain the link diagram
of Fig.~7b, in which three self-energy insertions spectate, whilst
the other two necessarily lead to entanglements. The resulting
4-component 12-crossing link diagram of Fig.~7c then delivers the
4-braid 9-crossing factor knot $\sigma_1^3\sigma_2^3\sigma_3^3$,
corresponding to the level-9 factorized transcendental $\ze3^3$.
Fig.~8 shows how the 4-braid 11-crossing prime knot
$\eltft\equiv11_{353}$
can be generated when there are dressings on all 6 lines of the tetrahedron,
forcing even more entanglements. In Fig.~8b, only two of the
6 self-energy insertions spectate, while the remaining 4 yield the
4-component 14-crossing link diagram of Fig.~8c. Three final skeinings
deliver the knot $11_{353}$. {}From this we conclude that the associated
knot number~\cite{EUL}
\begin{equation}
N_{3,5,3}=\ze{3,5,3}-\ze3\ze{5,3}-7\ze5\ze3^2\label{N353}
\end{equation}
will appear in the expansion of the master two-loop integral\Eq{I6},
though not in the ${\rm O}(1/N^3)$ anomalous dimensions.

By adding further self-energy insertions to Fig.~8, we can generate
a pair of 13-crossing 4-braids.
Their braid words and knot numbers are given in~\cite{EUL}. Here we are
content to stop at 12 crossings, with results summarized by Table~1.
Comparing these findings with the numbers of possible
non-zeta expansion coefficients in\Eq{non}, we arrive at the
following expectations:
\begin{enumerate}
\item[E8] The $\ep$-expansion of $\ep I(1-\ep)$ will contain the
double sum $\ze{5,3}$, associated with $8_{19}$, at ${\rm O}(\ep^8)$.
{}From\Eq{non} we see that is the sole level-8 irreducible in the
generic expansion of $\hyp$ series.
\item[E9] It will contain no new irreducible at ${\rm O}(\ep^9)$,
since the only  positive prime knot with crossing number 9 is the
$(9,2)$ torus knot, which delivers $\ze9$.
\item[E10] It will contain the double sum $\ze{7,3}$,
associated with $10_{124}$, at ${\rm O}(\ep^{10})$. The second level-10
non-zeta in the generic expansion of $\hyp$ series, indicated
by\Eq{non}, presumably involves
the product $\ze2\ze{5,3}$. Such a product cannot occur in MS counterterms,
since $\ze2$ is avoided in the $G$-scheme~\cite{CT}, which is equivalent to
the MS scheme for the calculation of anomalous dimensions.
\item[E11] It will not contain the triple sum $\ze{3,5,3}$, at
${\rm O}(\ep^{11})$. This term corresponds to the level-11 non-zeta term
in\Eq{lead}, but will not show up at ${\rm O}(1/N^3)$, since it requires more
than 3 dressed propagators to generate the entanglement of momentum flows
that skeins
to produce the positive hyperbolic\footnote{In~\cite{BKP},
$10_{139}$, $10_{152}$
and $11_{353}$ were wrongly called satellite knots. All three are, in fact,
hyperbolic.} knot $\eltft$, dubbed $11_{353}$ in~\cite{BKP}.
\item[E12] Two level-12 irreducible Euler sums are expected
in the expansion of the master two-loop diagram.
{}From the point of view of knot theory, it appears most natural to take these
as the alternating sums $N_{9,3}$ and $N_{7,5}$. If one wishes to stay
within the realm of non-alternating sums, the irreducibles must be
taken from different depths, with $\ze{9,3}$ appearing at depth 2,
and $\ze{4,4,2,2}$ at depth 4, which appears somewhat unnatural
for a pair of 3-braids with very similar structure. In any case,
no more than two new transcendentals should appear at 12 loops in the
anomalous dimensions of $(\phi^2)^2$ theory to order $1/N^3$ in the
large-$N$ limit.
\end{enumerate}
To check these expectations, we must tackle the generic
problem of expanding $\hyp$ series.

\subsection{Expansion to level 11 of any Saalsch\"utzian $\hyp$ series}

To extend expansion\Eq{lead} to level 11, we evaluated two instances
of\Eq{hdy}, which gave all the zeta-reducible coefficients at levels 7, 9,
and 11. Choosing a case that involved only double and triple sums,
we further reduced the remaining coefficients, save that of the invariant
$N(0,2,1,0)$,
to zetas and
the canonical knot-theoretic set of non-zetas up to level 11,
namely~\cite{BKP}
\begin{eqnarray}
\ze{5,3}=\sum_{m>n>0}\frac{1}{m^5n^3}&\approx&
0.037\,707\,672\,984\,847\,544\,011\,304\,782\,293\,659\,915
\label{F53}\,,\\
\ze{7,3}=\sum_{m>n>0}\frac{1}{m^7n^3}&\approx&
0.008\,419\,668\,503\,096\,332\,423\,968\,579\,714\,670\,651
\label{F73}\,,\\
\ze{3,5,3}=\sum_{l>m>n>0}\frac{1}{l^3m^5n^3}&\approx&
0.002\,630\,072\,587\,647\,467\,345\,248\,476\,381\,643\,627
\label{F353}\,,
\end{eqnarray}
associated with the knots $8_{19}$, $10_{124}$, and $11_{353}$.
Finally, we reduced the remaining
level-10
coefficient to this set and a single further non-zeta term:
\begin{equation}
\sum_{m>n_i>0}\frac{1}{m^6n_1n_2n_3n_4}
=\df{301509}{5600}\ze{10}
-\df{119}{4}\ze7\ze3
-\df{1413}{56}\ze5^2
-8\ze5\ze3\ze2
+\df{33}{2}\ze4\ze3^2
-\df{4}{5}\ze2\ze{5,3}
-\df{191}{56}\ze{7,3}\,,
\label{s10}
\end{equation}
whose value was obtained by high-precision evaluation and integer-relation
searching, enabled by Bailey's {\sc mpfun}~\cite{DHB} routines.
Excluding double
and triple sums, there are 42 sums of the form\Eq{s10}, with inverse powers
of $m$ and $n_i<m$ whose exponents sum to 10. Thanks to
{\sc mpfun}~\cite{DHB},
we discovered that {\em all\/} are reducible to the basis set that appears
in\Eq{s10}, which was the only case needed for our present field-theoretic
purposes. It is noteworthy that such sums do not suggest candidates for new
knot-transcendentals to associate with $10_{139}$ or $10_{152}$.

For the development
of the invariant expansion\Eq{lead}, we obtain
\begin{eqnarray}
\overline{W}(a,b,c,d)&=&
\sum
C(p_2,m_2,p_3,m_3)\,
N(p_2,m_2,p_3,m_3)+{\rm O}(\ep^{9})\,,
\label{wpc}\\
C(0,2,0,0)&=&
-\df{75}{16}\ze7
          +3\ze5\ze2
  -\df{3}{2}\ze4\ze3
\,,\nonumber\\
C(0,1,0,1)&=&\phantom{-}
\df{77}{20}\ze8
 -\df{9}{5}\ze{5,3}
\,,\nonumber\\
C(1,2,0,0)&=&
-\df{667}{48}\ze9
          +10\ze7\ze2
   -\df{7}{4}\ze6\ze3
  -\df{19}{4}\ze5\ze4
   -\df{1}{2}\ze3^3
\,,\nonumber\\
C(0,2,1,0)&=&
-\df{13023}{1120}\ze{10}
      +\df{91}{8}\ze7\ze3
   +\df{981}{112}\ze5^2
               -6\ze5\ze3\ze2
       +\df{5}{2}\ze4\ze3^2
    +\df{75}{112}\ze{7,3}
\,,\nonumber\\
C(1,1,0,1)&=&\phantom{-}
 \df{234}{35}\ze{10}
+\df{153}{28}\ze5^2
  -\df{27}{14}\ze{7,3}
\,,\nonumber\\
C(2,2,0,0)&=&
-\df{2293}{64}\ze{11}
  +\df{105}{4}\ze9\ze2
    -\df{1}{4}\ze8\ze3
   -\df{43}{8}\ze7\ze4
   -\df{77}{8}\ze6\ze5
   -\df{13}{4}\ze5\ze3^2
\,,\nonumber\\
C(0,4,0,0)&=&
-\df{177}{256}\ze{11}
  +\df{35}{36}\ze9\ze2
+\df{325}{288}\ze8\ze3
   +\df{11}{8}\ze7\ze4
  -\df{17}{16}\ze6\ze5
    -\df{7}{4}\ze5\ze3^2
    +\df{4}{9}\ze3^3\ze2
\,,\nonumber\\
C(0,1,1,1)&=&\phantom{-}
\df{583}{10}\ze{11}
 -\df{71}{5}\ze8\ze3
         -16\ze7\ze4
         -38\ze6\ze5
      +4\ze5\ze3^2
  +\df{9}{5}\ze3\ze{5,3}
  +\df{9}{5}\ze{3,5,3}
\,.\nonumber
\end{eqnarray}
which enables us to expand, to level 11, any Saalsch\"utzian $\hyp$
series\Eq{fis} whose parameters differ from integers by multiples of
$\ep$, and hence to expand the master two-loop diagram\Eq{inv} to this
order, thanks to the reduction of the case\Eq{I4} to a pair of series
in\Eq{I4is}, each of which is given by a $\hyp$ series in\Eq{sis}.

To expand\Eq{inv} in $2\mu=4-2\ep=\frac23\sum_n\al{n}$ dimensions,
one constructs $S_6\times Z_2$ invariants
from $N_k\equiv\sum_n\Delta_n^k$, where
$\Delta_1\equiv\frac16(\al1+3\al2-\al3+\al4-3\al5-\al6)$, with $\Delta_2$ to
$\Delta_6$ obtained by cyclic permutation of subscripts. In 4 dimensions,
we obtain
\begin{eqnarray}
\left.\overline{I}_6\right|_{\ep=0}&=&
6\ze3+\df{15}{2}N_2\ze5
+\df{63}{32}\left[5N_2^2-4N_4\right]\ze7
+\left[\df{3}{4}N_2^3-\df{9}{2}N_4N_2+6N_6-N_3^2\right]\ze3^3\nonumber\\
&+&\left[\df{85}{64}(\df{35}{3}N_2^3-28N_4N_2+\df{64}{3}N_6)
-\df{83}{9}N_3^2\right]\ze9
\nonumber\\
&+&\left[\df{45}{16}(N_2^4-6N_4N_2^2+8N_6N_2)-\df{9}{4}N_3^2N_2
-\df{18}{5}N_5N_3\right]\ze5\ze3^2\nonumber\\
&+&\left[\df{1023}{1024}(\df{151}{6}N_2^4-92N_4N_2^2+\df{256}{3}N_6N_2
+8N_4^2-\df{224}{9}N_3^2N_2)-\df{33}{8}N_5N_3\right]\ze{11}
\nonumber\\
&+&\left[\df{9}{5}N_5N_3-\df{3}{4}N_3^2N_2\right]
\left\{\df35(\ze{3,5,3}-\ze3\ze{5,3})+\df{69}{40}\ze{11}-\ze5\ze3^2\right\}
+{\rm O}(\Delta_n^{10})\,,
\label{I64}
\end{eqnarray}
where the final brace is the constant called $K_{353}$ in~\cite{BKP},
and was found in all those subdivergence-free 7-loop $\phi^4$
counterterms whose skeinings produce the knot $11_{353}$.
The even-level transcendentals $\ze{5,3}$
and $\ze{7,3}$ occur at levels 8 and 10 in the $\ep$-expansion. Note, however,
that only odd levels are encountered in\Eq{I64}, at $\ep=0$.
It follows that a counterterm producing the knot $8_{19}$
or $10_{124}$ {\em cannot\/} be reduced to a two-loop two-point form
merely by analytic regularization~\cite{AR} in 4 dimensions,
which is sufficient for the 6-loop zig-zag~\cite{5LB,BKP}
counterterm, $168\ze9$,
of $\phi^4$-theory.
This no-go theorem may help to prevent fruitless searches for
analytically regularized evaluations of counterterms that skein to
prime 3-braids.

\subsection{Expansion to level 11 of the large-$N$ integral $I(\mu)$}

The particular case of\Eq{I4is} required for $\eta$ at ${\rm O}(1/N^3)$
is given in\Eq{Pi}.
Near 4 and 2 dimensions, we obtain
\begin{eqnarray}
\Pi(2-\ep,\Delta)&=&\frac{2}{1-2\ep}\,
\frac{\Gamma(1+\Delta-\ep)\Gamma(1-\Delta+\ep)}
{\Gamma(1+\Delta-2\ep)\Gamma(1-\Delta)}\,
\overline{S}(\Delta-\ep,-\Delta)
\,,\label{im4}\\
\Pi(1-\ep,\Delta)&=&\frac{2(\Delta-3\ep)}{\Delta+\ep}\,
\frac{\Gamma(\Delta-\ep)\Gamma(-\Delta+\ep)}
{\Gamma(1+\Delta-2\ep)\Gamma(1-\Delta)}
+\frac{2\ep^2(1-2\ep)}{\Delta+\ep}\,
\Pi(2-\ep,\Delta)\,,\label{im2}
\end{eqnarray}
respectively, where the symmetric function
\begin{equation}
\overline{S}(a,b)\equiv S(a,b,a+b,0)/ab=3\ze3+{\rm O}(a,b)
\label{sb}
\end{equation}
is reducible if any element of $\{a,b,a+b,a+2b,2a+b\}$ vanishes, since
\begin{eqnarray}
2a\overline{S}(a,0)&=&
3\psi^\prime(1)-3\psi^\prime(1+a)
\,,\label{bz}\\
4a\overline{S}(2a,-2a)&=&
2\psi^\prime(1-2a)+\psi^\prime(1+a)-2\psi^\prime(1+2a)-\psi^\prime(1-a)
\,,\label{bm}\\
\frac{1+2a^3\overline{S}(a,-2a)}{\cos\pi a}&=&
\frac{\Gamma(1+a)\Gamma(1-3a)}{\Gamma(1-2a)}
\,.\label{b2}
\end{eqnarray}
Note that the coefficient of the last term in\Eq{im2} is ${\rm O}(\ep)$,
when $\Delta\sim\ep$. Thus level-$n$ terms, occurring
at $n$ loops in 4-dimensional theories, occur at $n+1$ loops
in 2-dimensional theories.

The special cases\Eqqq{bz}{bm}{b2} determine the expansion of\Eq{sb} to
level 7, giving
\begin{equation}
\ep\,I(1-\ep)
=\df23
+\df23\ze3\ep^3
+\ze4\ep^4
+\df{13}{3}\ze5\ep^5
+\left[\df{55}{6}\ze6+\df{11}{3}\ze3^2\right]\ep^6
+\left[\df{95}{4}\ze7+11\ze4\ze3\right]\ep^7+\sum_{n\geq8}X_n\ep^n.
\label{xi7}
\end{equation}
The first term was obtained in~\cite{Vas3}, the second in~\cite{BW},
the third in~\cite{et3,Vas4}, and the fourth in~\cite{et3}.
The zeta terms at levels 6 and 7 are new.
At level 8 we expect to encounter the first non-zeta term\Eq{F53},
corresponding to the knot $(4,3)=8_{19}$.

Using the general expansion\Eq{wpc}, we obtain the coefficients at
levels 8 to 11:
\begin{eqnarray}
X_8&=&\df{797}{15}\ze8+\df{74}{3}\ze5\ze3+\df{18}{5}\ze{5,3}\,,
\nonumber\\
X_9&=&\df{227}{2}\ze9+\df{130}{3}\ze6\ze3+10\ze5\ze4-\df{22}{3}\ze3^3\,,
\nonumber\\
X_{10}&=&\df{5953}{28}\ze{10}+\df{165}{2}\ze7\ze3+\df{836}{21}\ze5^2
-33\ze4\ze3^2+\df{54}{7}\ze{7,3}\,,
\nonumber\\
X_{11}&=&\df{5875}{12}\ze{11}+\df{3287}{30}\ze8\ze3-\df{153}{4}\ze7\ze4
+\df{200}{3}\ze6\ze5-64\ze5\ze3^2-\df{36}{5}\ze3\ze{5,3}\,,
\label{x11}
\end{eqnarray}
with $\ze{3,5,3}$ conspicuous by its {\em absence\/} at level 11,
as anticipated.

Referring back to the expectations E8 to E11 of Section~4.2, one sees that
knot theory is indeed a good guide, up to level 11.
To progress to level 12, and beyond, much more analysis was needed.
In fact, we eventually succeeded in reducing the $\hyp$ series
that occurs at ${\rm O}(1/N^3)$ to an elementary double sum, for all values
of $d=2-2\ep$.

\subsection{Reduction of $I(\mu)$ to a double sum, for all orders}

For all $|\ep|<1$, we define
\begin{equation}
S_\pm(\ep)\equiv\sum_{m>n>0}\frac{\ep^3}{(m+\ep)^2(n-\ep)}
\pm(\ep\to-\ep)\,{}=\sum_{r,s>0}(r-1)\ze{r,s}\left[(-1)^r\pm(-1)^s\right]
\ep^{r+s}\,,
\label{spm}
\end{equation}
where $\ze{r,s}=\sum_{m>n>0}m^{-r}n^{-s}$,
and $S_-(\ep)$, involving odd (and hence zeta-reducible) double sums,
may be reduced to products of polygammas,
by solving its recurrence relation. We find that
\begin{equation}
S_-(\ep)=\df12(\pz2+\pz{-2}-\pz1-\pz{-1})(\pp1+\pp{-1}+1)
+\df34(\pp1-\pp{-1})+(\pp{2}-\pp{-2})(\pz{1}-\pz{-1}-1)\,,
\label{sm}
\end{equation}
where we use the shorthand
$\psi_p^{(n)}\equiv\left[\partial/\partial p\right]^{n+1}
\ln\Gamma(1+p\,\ep)=\ep^{n+1}\psi^{(n)}(1+p\,\ep)$
for the polygamma functions.  On the other hand, $S_+(\ep)$
is not reducible; its
expansion in $\ep$ involves even-level double sums,
$\ze{r,s}$ with $(-1)^r=(-1)^s$,
which cannot all be expressed as linear combinations of products
of zetas. In fact~\cite{BBG} the number of independent irreducible
non-alternating double sums, at level $r+s=2n$,
is the integer part of $(n-1)/3$.
In particular, one cannot
further reduce the expansion
\begin{eqnarray}
S_+(\ep)&=&\df12\ze4\ep^4+\left[6\ze3^2-\df{47}{6}\ze6\right]\ep^6
+\left[-\df{36}{5}\ze{5,3}+24\ze5\ze3-\df{863}{30}\ze8\right]\ep^8\nonumber\\
&&{}+\left[-\df{108}{7}\ze{7,3}+\df{96}{7}\ze5^2+36\ze7\ze3
-\df{4019}{70}\ze{10}\right]\ep^{10}+{\rm O}(\ep^{12})\,,\label{sp}
\end{eqnarray}
which will contain one more non-zeta term at level 12; two at each of levels
14, 16, and 18; three at levels 20, 22 and 24; and so on.
Since no four-fold sum occurs in expansion\Eq{spm}, no
alternating double sum occurs, which is why only one of the
two knot-numbers occurs at level 12. In the generic case of expanding
the master two-loop diagram, both are expected to occur.

We now show that\Eq{sp} is the {\em sole\/} source of non-zetas
in the expansion
of $I(1-\ep)$. Consider first the $\hyp$ series
$F(2a+b,-b,-a-b,-b)$, which gives\Eq{sb}, via\Eq{sis}.
Its expansion to ${\rm O}(a^2)$ is required, to obtain $I(\mu)$.
This entails two distinct series
in the expansion of the wreath-product invariant\Eq{sub}, of the form
\begin{equation}
\overline{W}(2a+b,-b,-a-b,-b)=
\sum_{n=0}^\infty c_n N(n,2,0,0) +
\sum_{n=0}^\infty d_n N(n,1,0,1)+{\rm O}(a^3)\,.\label{wab}
\end{equation}
The coefficients $c_n$ of the first series are those giving the
${\rm O}(a^2)$ terms in the zeta-reducible case\Eq{hdy}; those of
the second series give the ${\rm O}(h k)$ terms in
$F(h+\ep,-k-\ep,-\ep,k+\ep)$, which can in turn be related
to the expansion of $S_+(\ep)$. Thus one may use wreath-product
invariance to relate $I(1-\ep)$ to $S_+(\ep)$, using Hardy's
result\Eq{hdy}. The algebra is rather demanding: for each of the
three series one must determine the zeta-reducible
terms of\Eqqq{fb}{wp}{sub}, which involves taking third
derivatives of a very large number of $\Gamma$ functions, generating
$\{\psi^{(n)}_p\,|\,n=0,1,2\,;\,p=0,\pm1,\pm2\}$ and their many products.
Some of these products are related by trigonometric identities,
involving the even-zeta parts of
\begin{equation}
\psi^{(n)}_p
=\frac12\left(\frac{\partial}{\partial p}\right)^{n+1}\!\!\left[
\ln\frac{\Gamma(1+p\,\ep)}{\Gamma(1-p\,\ep)}
+\ln\frac{p\,\pi\ep}{\sin p\,\pi\ep}\right]\,.
\label{trig}
\end{equation}
After considerable use of {\sc reduce}~\cite{RED}, we obtained
\begin{eqnarray}
&&\left(\ep\,I(1-\ep)+\pz1+\pz{-2}-\pz{-1}-\pz0\right)\left(1+\pp{-1}-\pp0
\right)-\df23\equiv
\nonumber\\
&&\quad E(\ep)
=-\df12S_+(\ep)+\df1{16}\left(2\pz{-1}-2\pz1+6\pp1+10\pp{-1}
-12\pp0+\pg1+\df73(\pg{-1}+2\pg0)\right)\nonumber\\
&&\qquad{}+\df14(\pz2-\pz1)(1+\pp1+\pp{-1}-6\pp0)
\,+\,\df14(\pz{-2}-\pz{-1})(1+\pp1+\pp{-1}+2\pp0)\,,
\label{Eis}
\end{eqnarray}
and verified that the expansion coefficients of\Eqqq{xi7}{x11}{sp} satisfy
this remarkable identity to ${\rm O}(\ep^{11})$. It is hard to imagine
how it might have been obtained without systematizing the 72 wreath-product
transformations of matrix\Eq{mat}. We conclude that the group-theory of
Saalsch\"utzian $\hyp$ series is useful for obtaining all-order results;
not just for perturbative expansions.
Comprehensive checks are provided by
the recurrence relation
\begin{equation}
\frac{E(\ep)+\frac23}{\ep^3}=
\frac{E(\ep+1)}{(\ep+1)^3}+\frac{\psi(-\ep)-\psi(1)}{2\ep(1+2\ep)}\,,
\label{erec}
\end{equation}
which follows from\Eqq{im4}{im2}, and is satisfied by\Eqq{spm}{Eis}, and by
evaluating\Eq{Eis} at $\ep=-\frac12$, where
the zeta-reducibility of\Eq{spm} yields a finite 3-dimensional
result:
\begin{equation}
S_{+}\!\left(\pm\df12\right)=-\df78\ze3+\df98\ze2-\df34
\quad\Rightarrow\quad I\!\left(\df32\right)=2\ln2-\frac{7\ze3}{2\ze2}\,,
\label{I32}
\end{equation}
in agreement with~\cite{Vas3}.

Thus we conclude the comparison with knot theory with the finding that
only one new non-zeta level-12 transcendental occurs in the critical
exponent $\eta$ at ${\rm O}(1/N^3)$. It may be taken as the double sum
$\ze{9,3}$, or indeed any member of $\{\ze{12-s,s}\mid s=2\ldots5\}$,
since this set contains only one~\cite{BBG} independent non-zeta term.
We incline to associate it with
$\sigma_1^{}\sigma_2^7 \sigma_1^{}\sigma_2^3$
though there is at present no strong reason
for preferring this knot to
$\sigma_1^{}\sigma_2^5 \sigma_1^{}\sigma_2^5$, or some linear combination
of their knot numbers.
The braid words of the remaining five 12-crossings knots
in~Table~1 make them unlikely candidates to be associated
exclusively with double sums.

\section{Computation of anomalous dimensions}

We now use the all-order result\Eq{Eis} to compute critical exponents
and anomalous dimensions in four field theories: the bosonic $\sigma$-model
in 2-dimensions; its 4-dimensional cousin, which is $\phi^4$ theory;
its fermionic cousin, which is the Gross-Neveu model; and its
supersymmetric extension, which has the attractive feature
that at $L$ loops all terms have transcendentality level $L-1$.
In this last case we give analytic results to 12 loop; space does not permit
us to write the explicit perturbation expansions in the other cases,
though they may be obtained from the results that we give.
Using\Eq{spm}, it is possible to develop the expansions numerically,
to any desired accuracy and number of loops. Hence we are able to investigate
the behaviour of Pad\'e resummations, up to 24 loops, and apply them
outside the domain of convergence of the original perturbation series.

\subsection{Bosonic $\sigma$-model and $\phi^4$-theory}

The critical exponent $\eta=\sum_k\eta_k/N^k$ of the bosonic $\sigma$-model,
in $2\mu=2-2\ep$ dimensions, is given to ${\rm O}(1/N^3)$ by~\cite{Vas3}
\begin{eqnarray}
\eta_1&=&-2\ep A\,\frac{R_0}{1-\ep}\,;\qquad
A\equiv\frac{\Gamma(1-2\ep)}{\Gamma(1+\ep)\left[\Gamma(1-\ep)\right]^3}\,,
\label{eb1}\\
\eta_2&=&-4\ep A^2\,\frac{R_1+R_2B}{(1-\ep)^2}\,;\qquad
B^{(n)}\equiv
\psi^{(n)}_1+(-1)^n\psi^{(n)}_{-2}-(-1)^n\psi^{(n)}_{-1}-\psi^{(n)}_0\,,
\label{eb2}\\
\eta_3&=&-8\ep A^3\,
\frac{R_3+R_4B+R_5B^2+R_6C+R_7B^\prime+R_8B C+R_9D+R_{10}E}{(1-\ep)^3}\,;
\label{eb3}\\
&&C\equiv\pp{-1}-\pp0\,;\,\,
D\equiv3B B^\prime-B^{\prime\prime}-B^3\,;\,\,
E\equiv(\ep I(1-\ep)+B)(1+C)-\df23\,,\label{cde}
\end{eqnarray}
where $\{B,B^\prime,B^{\prime\prime},C,D\}$ entail the polygammas\Eq{trig},
$E$ has been reduced to $S_+(\ep)$ in\Eq{Eis},
and $R_n$ are rational functions of $\ep=1-\mu$,
which we obtain from~\cite{Vas2,Vas3} as:
\begin{eqnarray}&&
R_0=1+\ep
    \,;\quad
R_1=(1-\ep^2+2\ep^3-4\ep^4)/(1-\ep)
    \,;\quad
R_2=-(1-\ep+2\ep^2)(1+\ep)
    \,;\nonumber\\&&
R_3=(1-\ep-\ep^2+5\ep^3-9\ep^4+\ep^5-8\ep^6+25\ep^7-11\ep^8+2\ep^9)/(1-\ep)^2
    \,;\nonumber\\&&
R_4=-(4-6\ep+4\ep^2+13\ep^3-11\ep^4-11\ep^5-7\ep^6+2\ep^7)/(1-\ep)
    \,;\nonumber\\&&
R_5=-2\ep(3-4\ep-7\ep^2+2\ep^3)
    \,;\qquad\quad\,\,\,
R_6=-\df14(3+\ep-25\ep^2+49\ep^3-18\ep^4-2\ep^5)
    \,;\nonumber\\&&
R_7=1+5\ep-5\ep^2-13\ep^3+4\ep^4+4\ep^5
    \,;\,\,
R_8=-(1+4\ep+7\ep^2)(1-\ep)^2
    \,;\nonumber\\&&
R_9=-\df23(1+2\ep)^2(1-\ep)^2
    \,;\qquad\qquad\quad\,\,
R_{10}=\df32(1+4\ep)(1-\ep)^2(1+\ep)
    \,.\label{rn}
\end{eqnarray}
{}From~\cite{Vas2}, we obtain
\begin{equation}
\lambda_1=\frac{2\ep^2(1-2\ep)A}{(1-\ep)}\,;\quad
\lambda_2=\frac{2\ep(S_1+S_2B+S_3(B^2-B^\prime)+S_4C)A^2}
{(1+\ep)(1-\ep)^3}\,,\label{la2}
\end{equation}
for $\lambda=1/2\nu=\sum_k \lambda_k/N^k$, to ${\rm O}(1/N^2)$,
with polynomial coefficients
\begin{eqnarray}&&
S_1=2\ep(1-2\ep)(1-\ep+4\ep^2-7\ep^3-2\ep^4+3\ep^5)
\,;\nonumber\\&&
S_2=-2(1-\ep)(2+5\ep-6\ep^2-9\ep^3+4\ep^5)
\,;\nonumber\\&&
S_3=2(1-\ep)^3(1+2\ep)^2\,;\quad
S_4=-3(1-\ep)^3(1+5\ep+8\ep^2)\,.\label{sla2}
\end{eqnarray}
Extending the techniques of~\cite{Vas2} to include
corrections to the asymptotic scaling forms of the propagators due to
insertion of an operator with dimension $(\mu-2)$, we obtained
\begin{equation}
\omega_1=\frac{4\ep(1-2\ep)^2(1+\ep)A}{(1-\ep)}\,;\quad
\omega_2=-\frac{4\ep T_0A+4\ep(T_1+T_2B+T_3(B^2-B^\prime)+T_4C)A^2}
{(2+\ep)^2(1+\ep)^2(1-\ep)^3}\,,\label{om2}
\end{equation}
with polynomial coefficients
\begin{eqnarray}&&
T_0=-16(1+2\ep)^2(1-\ep)^3
\,;\quad
T_1=8+30\ep-94\ep^2-49\ep^3+372\ep^4-3\ep^5
\nonumber\\&&\qquad{}
-612\ep^6-271\ep^7+556\ep^8+609\ep^9-70\ep^{10}-260\ep^{11}-72\ep^{12}
\,;\nonumber\\&&
T_2=-2(1-\ep)(20+118\ep+177\ep^2-36\ep^3-298\ep^4-278\ep^5-131\ep^6+52\ep^7
\nonumber\\&&\qquad{}
+136\ep^8+80\ep^9+16\ep^{10})
\,;\quad
T_3=8(2+\ep)(1+\ep)(1+2\ep)^2(1-\ep)^3(1+3\ep+\ep^2)
\,;\nonumber\\&&
T_4=-3(2+\ep)(1+\ep)(1-\ep)^3(2+19\ep+74\ep^2+89\ep^3+28\ep^4+4\ep^5)
\,,\label{tom2}
\end{eqnarray}
also in $d=2-2\ep$ dimensions.

Working to 5 loops at ${\rm O}(1/N^3)$ we obtain the $\ep$-expansion
\begin{equation}
\eta_3=-8\ep-32\ep^2-72\ep^3-4(40+27\ze3)\ep^4-18(16+4\ze3+9\ze4)\ep^5
+{\rm O}(\ep^6)\,,\label{et32}
\end{equation}
for the $\sigma$-model in $2-2\ep$ dimensions. Using\Eq{erec}
to shift the dimensionality by two units, we obtain the $\ep$-expansions
\begin{eqnarray}
\eta_3&=&
320\ep^2
-1984\ep^3
+4(683-240\ze3)\ep^4
\nonumber\\&&{}
+2(343+4720\ze3-720\ze4+1280\ze5)\ep^5
+{\rm O}(\ep^6)\label{et34}\,,\\
\lambda_2&=&
-48\ep
+242\ep^2
+(-283+240\ze3)\ep^3+
(59-1328\ze3+360\ze4-640\ze5)\ep^4
\nonumber\\&&{}
+8(-22+279\ze3-32\ze3^2-249\ze4+620\ze5-200\ze6)\ep^5
+{\rm O}(\ep^6)\label{la34}\,,\\
\omega_2&=&
408\ep^2
+4(-259+240\ze3)\ep^3
+96(-18-63\ze3+15\ze4-40\ze5)\ep^4
\nonumber\\&&{}
+2(1591+6432\ze3-1152\ze3^2-4536\ze4+19520\ze5-4800\ze6)\ep^5
+{\rm O}(\ep^6)\,,\label{om34}
\end{eqnarray}
for $(\phi^2)^2$ theory, in $4-2\ep$ dimensions.
Results\Eqqq{et34}{la34}{om34} agree with~\cite{PHI}.

\subsection{Gross-Neveu model}

For the Gross-Neveu model, with a four-fermion interaction,
one has merely to replace,
in\Eqqq{eb1}{eb2}{eb3}, the rational factors $R_n$ by~\cite{Der}
\begin{eqnarray}&&
\tilde{R}_0=-\ep
    \,;\quad
\tilde{R}_1=-\df12\ep(3-4\ep)(1-2\ep)/(1-\ep)
    \,;\quad
\tilde{R}_2=\ep(1-2\ep)
    \,;\nonumber\\&&
\tilde{R}_3=-\df12\ep(4-18\ep+24\ep^2-2\ep^3-14\ep^4+6\ep^5-\ep^6)(1-2\ep)
           /(1-\ep)^2
    \,;\nonumber\\&&
\tilde{R}_4=\df12\ep(8-26\ep+19\ep^2+3\ep^3-\ep^4)(1-2\ep)/(1-\ep)
    \,;\quad
\tilde{R}_5=-3\tilde{R}_7=-\df32\ep(1-2\ep)^2
    \,;\nonumber\\&&
\tilde{R}_6=-\df14\ep(15-32\ep+17\ep^2-\ep^3)
    \,;\quad
\tilde{R}_8=\tilde{R}_{10}=\df32\ep(1-\ep)^2
    \,;\quad
\tilde{R}_9=0
    \,.\label{rtn}
\end{eqnarray}

\subsection{Supersymmetric $\sigma$-model}

Beyond 5 loops, the bosonic and fermionic
expansions become lengthy to write,
because of the mixing of transcendentals of different levels.
To exemplify the generic transcendentality structure we consider,
instead, the
supersymmetric $N$-component $\sigma$-model of~\cite{et3,et2,lm2},
in $2\mu=2-2\ep$ dimensions, which is free of such level mixing.
We expand the anomalous dimensions, $\beta(g)$ and $\gamma(g)$,
of the coupling and field, in powers of $1/N$, as follows:
\begin{eqnarray}
\frac{2\ep g+\beta(g)}{(2-N)g^2}&\equiv&\overline\beta(g)
=1+(N-3)\sum_{k=2}^\infty(-N)^{-k-1}
\beta_k(-\df{N}{2}g)\,,\label{bb}\\
\frac{\gamma(g)}{(N-1)g}&\equiv&\overline\gamma(g)
=1+(N-2)\sum_{k=1}^\infty (-N)^{-k-1}
\gamma_k(-\df{N}{2}g)\,,\label{gb}
\end{eqnarray}
since multi-loop terms in $\beta$ and $\gamma$
vanish for $N=3$ and $N=2$, respectively.
At ${\rm O}(1/N)$ there is no contribution to $\beta$, while
\begin{equation}
\gamma_1(\ep)=A-1=2\ze3\ep^3+3\ze4\ep^4+6\ze5\ep^5+{\rm O}(\ep^6)\,.
\label{gam1}
\end{equation}
The critical coupling is determined by $\beta(g_c)=0$, which gives
\begin{equation}
-\frac{N}{2}g_c=\ep\left[1+\frac{2}{N}
+\frac{4+\beta_2(\ep)}{N^2}+{\rm O}(1/N^3)\right]\,,
\label{gc}
\end{equation}
where the ${\rm O}(1/N^2)$ term in the critical exponent
$\beta^\prime(g_c)\equiv-2\lambda$ gives~\cite{lm2}
\begin{equation}
\ep\beta_2^\prime(\ep)=-2A^2(4B-2B^2+3C+2B^\prime)
=36\ze3\ep^3+54\ze4\ep^4+232\ze5\ep^5+{\rm O}(\ep^6)\,,
\label{bet2}
\end{equation}
in terms of the polygammas of\Eqq{eb2}{cde}.
The results of~\cite{et3,et2} give
\begin{eqnarray}
\frac{(N-1)\overline\gamma(g_c)}{(N-2)\overline\beta(g_c)}&=&1
-\frac{\eta}{2\ep}\,=\,1+A/N+2(1-B)A^2/N^2\nonumber\\
&&{}+4(1-4B-\df34C+B^\prime-B C-\df23D+\df32E)A^3/N^3+{\rm O}(1/N^4)\,,
\label{eta}
\end{eqnarray}
where the coefficients of the polygammas
are obtained by the simple device~\cite{et3}
of setting $\ep=0$ in the rational functions of\Eq{rn}.
{}From\Eq{eta} we obtain the ${\rm O}(1/N^2)$ terms in $\gamma$:
\begin{equation}
\gamma_2(\ep)=\beta_2(\ep)-\left(2[\gamma_1(\ep)-\ep\,\rd/\rd\ep]
+5\right)\gamma_1(\ep)+2A^2B
=6\ze3\ep^3+\df{21}{2}\ze4\ep^4+\df{222}{5}\ze5\ep^5+{\rm O}(\ep^6)\,,
\label{gam2}
\end{equation}
so that $\gamma_2(\ep)=3\gamma_1(\ep)+{\rm O}(\ep^4)$, which means that
the 4-loop term in $\gamma$ vanishes for $N=3$.

The non-zeta transcendentals $\ze{5,3}$ and $\ze{7,3}$
are found at 9 and 11 loops, respectively, in the expansion
of $\beta_3-\gamma_3$.
Using\Eqq{sp}{Eis}
in\Eq{eta}, we obtain
\begin{eqnarray}
\beta_3(\ep)-\gamma_3(\ep)&=&
 \df{45}{2}\ze4\ep^4
+\df{1328}{5}\ze5\ep^5
+\df53\left[212\ze3^2+389\ze6\right]\ep^6
+\df{1}{14}\left[17232\ze4\ze3+30629\ze7\right]\ep^7\nonumber\\
&+&\!\!\!\df25\left[-54\ze{5,3}+12613\ze5\ze3+16633\ze8\right]\ep^8\nonumber\\
&+&\!\!\!\df{1}{9}\left[16532\ze3^3+73494\ze5\ze4+97730\ze6\ze3
          +129669\ze9\right]\ep^9\nonumber\\
&+&\!\!\!\df{1}{35}\!\left[-1620\ze{7,3}+309797\ze4\ze3^2+449566\ze5^2
           +1033435\ze7\ze3+1896379\ze{10}\right]\ep^{10}\nonumber\\
&+&\!\!\!\!\df{1}{770}\!\left[-99792\ze3\ze{5,3}+22188152\ze5\ze3^2
+40049660\ze6\ze5
\right.\nonumber\\&&\left.\quad{}
                         +35783955\ze7\ze4+64668044\ze8\ze3
                         +65848405\ze{11}\right]\ep^{11}+{\rm O}(\ep^{12})\,,
\label{bg3}
\end{eqnarray}
to 12 loops in the perturbation expansion, with a coupling $g=-2N\ep$.
Note that we cannot separate the non-zeta contributions
of $\beta_3$ and $\gamma_3$ to $\eta_3$, without also knowing $\lambda_3$.
If we parametrize
the unknown terms of the 6-loop anomalous dimension by $\delta_{1,2,3}$ in
\begin{eqnarray}
\gamma(g)&=&(N-1)g-\df14\ze3(N-1)(N-2)(N-3)g^4\nonumber\\
&+&\df{3}{32}\ze4(N-1)(N-2)\left[(N-3)(2N-1)+\delta_1\right]g^5\label{g6}\\
&-&\df{3}{80}\ze5(N-1)(N-2)\left[(N-3)(5N^2-22(N-1)+\delta_2)
+\delta_3\right]g^6
+{\rm O}(g^7)\,,\nonumber
\end{eqnarray}
then\Eqq{bet2}{bg3} give the 6-loop $\beta$-function as
\begin{eqnarray}
\beta(g)/g&=&2\mu-2-(N-2)g-\df32\ze3(N-2)(N-3)g^4\nonumber\\
&+&\df{27}{32}\ze4(N-2)(N-3)\left[N-(2+\df19\delta_1)\right]g^5\label{b6}\\
&-&\df{29}{20}\ze5(N-2)(N-3)\left[N^2-(8+\df{3}{116}\delta_2)N
+\delta_4\right]g^6
+{\rm O}(g^7)\,,
\nonumber
\end{eqnarray}
where $\delta_4$ involves $\lambda_4$. The vanishing of the 4-loop term in
$\gamma$ at $N=3$, like all terms beyond one loop in $\beta$,
suggests $\delta_1=0$, giving simple factors\footnote{JAG regrets errors
in previous work. To correct these:
in Eq~(14) of~\cite{et3} replace $\frac23$ by $\frac13$;
hence in Eq~(15) replace $(2N-3)(N-2)$ by $(2N-1)(N-3)$;
in Eq~(5.11) of~\cite{lm2} delete the $\ze3$ term at 6 loops and
change the sign of the $\ze3^2$ term at 7 loops; hence in Eq~(5.15)
delete the $\ze3$ term at 6 loops.}
of $N-3$ and $N-2$
in the 5-loop terms of\Eqq{g6}{b6}, respectively.
Similarly, $\delta_{3}=0$ gives a factor of $N-3$
at 6 loops in $\gamma$, and $\delta_2=0$ simplifies $\beta$.
In such a case, the 6-loop term
in $\beta_3-\gamma_3$ comes from that in $8\beta_2-4(\gamma_2-3\gamma_1)$,
with\Eqqq{gam1}{bet2}{gam2} giving
$8(232/5)-4(222/5-18)=1328/5$ in\Eq{bg3}.

\subsection{Pad\'e approximation of $\ep$-expansions}

Suppose that one knows the expansions of anomalous dimensions to some
order in the perturbation theory of the 2-dimensional field theory. How
accurately can one estimate 3-dimensional exponents?

{}From $L$-loop perturbation theory one may clearly
construct $L$ terms of the $\ep$-expansion
of the ${\rm O}(1/N^3)$ term $\eta_3$. Setting $\ep=-\frac12$ will {\em not}
give a reasonable estimate of the 3-dimensional value, since there are
singularities at $2\mu=1$ that limit the convergence of the series
to $|\ep|<\frac12$. These are severe: in the supersymmetric case,
the coefficient of $\ep^n$ increases like $n^5 2^n$,
on account of a $(1-2\ep)^{-6}$ singularity
in the $A^3D$ term of\Eq{eta}. It is for this reason that the
coefficients of\Eq{bg3} are so large.

In such a situation, one may proceed by calculating Pad\'e approximants of
the form
\begin{equation}
\eta_3\approx\frac{\sum_{n=1}^{L\!-\!M}a_n\ep^n}{1+\sum_{m=1}^M b_m\ep^m}
\equiv[L\!-\!M\backslash M]\,,
\label{pad}
\end{equation}
with coefficients chosen to fit the $L$-loop $\ep$-expansion.
Extending\Eq{sp}, we generate
\begtab
  \tline{12}{  50.6}{1.618343}{1.816265}{1.739471}{1.501598}
  \tline{16}{ 190.6}{1.737426}{1.737543}{1.722441}{1.734508}
  \tline{20}{ 520.2}{1.733445}{1.728054}{1.728646}{1.728264}
  \tline{24}{1178.6}{1.733935}{1.728344}{1.728335}{1.728337}
\tabend{tss2}
as approximants to the supersymmetric 3-dimensional result~\cite{et3}
\begin{equation}
\eta_1=\frac{8}{\pi^2}\,;\quad\eta_2=\eta_1^2\,;\quad\eta_3=
\left[2-(3\ln2+1)\ze2+\df{21}{4}\ze3\right]\eta_1^3
\approx1.728337\,.
\label{sdim3}
\end{equation}
Direct summation, corresponding to $[L\backslash0]$, is
clearly not an option.
Note that at $L=12$ loops the second significant figure is in doubt;
it is sobering to realize how deep into the $\ep$-expansion one must
go to get close to the non-perturbative value.

For the Gross-Neveu model, the results~\cite{Der} of\Eq{rtn} give the
3-dimensional values\footnote{In~Eq~(4.7) of~\cite{gn2}, 244 should be
replaced by 224.
In~Eq~(5.1) of~\cite{gn3}, the final term should be divided by 2; hence
167 should be replaced by 653 in~Eq~(5.2).}
\begin{equation}
\eta_1=\frac{8}{3\pi^2}\,;\quad\eta_2=\df{28}{3}\eta_1^2\,;\quad\eta_3=
-\left[\df{653}{18}-(27\ln2+\df{47}{4})\ze2+\df{189}{4}\ze3\right]\eta_1^3
\approx-0.847408\,,
\label{gdim3}
\end{equation}
to be compared with the Pad\'e table
\begtab
  \tline{12}{12.86}{-1.768424}{-0.816904}{-1.050237}{ 0.715507}
  \tline{16}{26.58}{-0.844409}{-0.850763}{-0.839889}{-0.841839}
  \tline{20}{46.98}{-0.847379}{-0.846989}{-0.847284}{-0.847540}
  \tline{24}{75.42}{-0.847306}{-0.847406}{-0.847409}{-0.847405}
\tabend{tgn2}
which is similarly slow to settle down.

Continuing the bosonic $\ep$-expansion of\Eq{et32}, we obtain the Pad\'e
table
\begtab
  \tline{12}{ 233.3}{-1.514539}{-1.725373}{-2.423265}{-0.770594}
  \tline{16}{ 991.4}{-1.839365}{-2.008766}{-1.797160}{-1.870298}
  \tline{20}{2888.6}{-1.847849}{-1.890675}{-1.881383}{-1.881535}
  \tline{24}{6847.2}{-1.842803}{-1.881245}{-1.881240}{-1.881215}
\tabend{tbs2}
to be compared with the exact 3-dimensional result~\cite{Vas3}
\begin{equation}
\eta_1=\frac{8}{3\pi^2}\,;\quad\eta_2=-\df83\eta_1^2\,;\quad\eta_3=
-\left[\df{797}{18}-(27\ln2-\df{61}{4})\ze2+\df{189}{4}\ze3\right]\eta_1^3
\approx-1.881235\,.
\label{bdim3}
\end{equation}
The convergent $\ep$-expansion
of\Eq{et34}, for $\phi^4$-theory in $4-2\ep$ dimensions, gives
\begtab
  \tline{12}{-3.205048}{-2.899329}{-1.449140}{-2.033950}{-2.675782}
  \tline{16}{-2.067376}{-1.861968}{-1.881627}{-1.885853}{-1.876600}
  \tline{20}{-1.880789}{-1.881162}{-1.881278}{-1.881218}{-1.881232}
  \tline{24}{-1.881231}{-1.881235}{-1.881234}{-1.881234}{-1.881235}
\tabend{tbs4}
for the same exponent. The convergence is almost as slow as for
the $\sigma$-model, in\Eq{tbs2}.

\section{Conclusions}

Our findings have consequences for field theory, number theory, and knot
theory.

{}From the point of view of field theory, we have reduced the
two-loop diagram of Fig.~2a, with up to three dressed lines,
and two adjacent lines free of dressings, to a pair of $\hyp$ series.
A particular case yields ${\rm O}(1/N^3)$ critical exponents in terms
of $\Gamma$ functions, their derivatives, and a single source of
non-zetas, which we have reduced, via\Eq{Eis}, to an elementary
double sum in\Eq{spm}, whose
$\ep$-expansion generates non-alternating Euler double sums.
The 12-loop analytical result\Eq{bg3} ensues in the supersymmetric
$\sigma$-model.
In other field
theories, perturbative expansions contain the same transcendentals but are
lengthier, due to level mixing. Numerical results are now obtainable to
any desired order and level of precision. Pad\'e resummation of
$\ep$-expansions, in four distinct field theories, to 24 loops,
reproduces analytical 3-dimensional results, albeit with painfully
slow convergence.

{}From the point of view of number theory, we have given a systematic
method to exploit the group theory of Saalsch\"utzian
$\hyp$ series, so as to obtain the expansion of\Eq{fis}
from~(\ref{fb}--\ref{root}), in terms of $\Gamma$
functions and the wreath-product invariant expansion\Eq{wpc},
whose non-zeta Taylor coefficients are enumerated in\Eq{non},
whilst the zeta-reducible ones are obtained from~(\ref{fab}--\ref{hdy}).
The non-zeta terms are irreducible Euler sums: at level 8, and again at
level 10, one encounters an irreducible double sum;
at level 11, an irreducible triple sum; and at level 12, two irreducibles,
which may be taken as a pair of alternating
double sums, or as a non-alternating double sum and a
non-alternating quadruple sum.

{}From the point of view of knot theory, all five irreducibles
with levels up to 12 are associated with positive knots more complex
than the $(2n-3,2)$ torus knot that produces $\ze{2n-3}$ in $n$-loop
counterterms. Table~1 shows the associations of the 3-braids $8_{19}$
and $10_{124}$ with alternating double sums of the form\Eq{nab},
and of the uniquely positive hyperbolic 11-crossing 4-braid with the
knot-number\Eq{N353}, which is first irreducible member of a class of
triple sums whose general form is given in~\cite{EUL}.
At 12-crossings, we associate the first two of the seven 3-braid knots of
Table~1 with linear combinations of the knot numbers $N_{9,3}$
and $N_{7,5}-\frac{\pi^{12}}{2^5 10!}$, though we have no method, at present,
to determine these linear combinations, since diagrams that entail double sums
give link diagrams whose skeinings generate both knots.
If only one knot is associated with the
non-alternating double sum of the ${\rm O}(1/N^3)$ results, then it is
probably $\fk{}{7}{}{3}$, whose Jones polynomial is simpler than that of
the other candidate, $\fk{}{5}{}{5}$.
However, there is no compelling reason to suppose
that the absence of alternating double sums at ${\rm O}(1/N^3)$
is a signal that only one member of the pair is entailed.
At higher order in $1/N$, we confidently expect that
alternating double sums will occur. More generally,
at crossing number $2m+4$, we associate
the $[m/2]$ knots, in\Eq{knots},
with linear combinations of the $[m/2]$
numbers, in\Eq{numbers}, modulo a multiple of $\pi^{2k+2l+4}$ that
is required to make $N_{2k+3,2l+1}$ a knot-number, for $k>l$.
Further arguments in favour of this association will be presented
in~\cite{BK14}, where knots with up to 14 crossings are studied.
Conspicuous among our findings is the impossibility
of generating the knots $10_{139}$ and $10_{152}$ from counterterms
obtained by arbitrary dressings of the tetrahedron whose skeleton
delivers $\ze3$. This strengthens our belief that their knot-numbers
entail transcendentals more complex than Euler sums, as is expected
from the fact that they have been generated, to date, only by the
most complex of 7-loop counterterms, with multiple sums
weighted by the squares of 6--j symbols~\cite{BKP}.

In conclusion: expansions of critical exponents and anomalous
dimensions at ${\rm O}(1/N^3)$ entail non-zeta terms because the
link diagrams that encode the intertwining of momenta in the
associated Feynman diagrams yield the knots\Eq{knots}.
The coefficients of the associated
transcendentals are now obtainable, via\Eq{Eis}, to arbitrarily high order,
thanks to application of the wreath-product transformations of~\cite{DJB}.
The knot theory of~\cite{DK1,DK2,BKP}, applied to this problem,
has led to discoveries in number theory~\cite{EUL}.

\noindent{\bf Acknowledgements}
We are grateful to the organizers and participants of the AIHEP workshop in
Pisa (April 95), the multiloop workshop in Aspen (August 95),
and the UKHEP Institute in Swansea (September 95),
which contributed greatly to our collaboration.
DJB thanks David Bailey for advice on using {\sc mpfun}~\cite{DHB},
Jon Borwein for correspondence on Euler sums~\cite{BBG}, and Tony Hearn
for enhancements to {\sc reduce}~\cite{RED}. DK acknowledges the support
of DFG and thanks Bob Delbourgo
for advice and encouragement in developing the connections between
knot theory and field theory.
DJB and DK also thank Bob Delbourgo for hospitality at the University
of Tasmania, which enabled completion of this work.

\newpage
\raggedright

\newpage

\begin{figure}[ht]
\epsfysize=13cm\epsffile{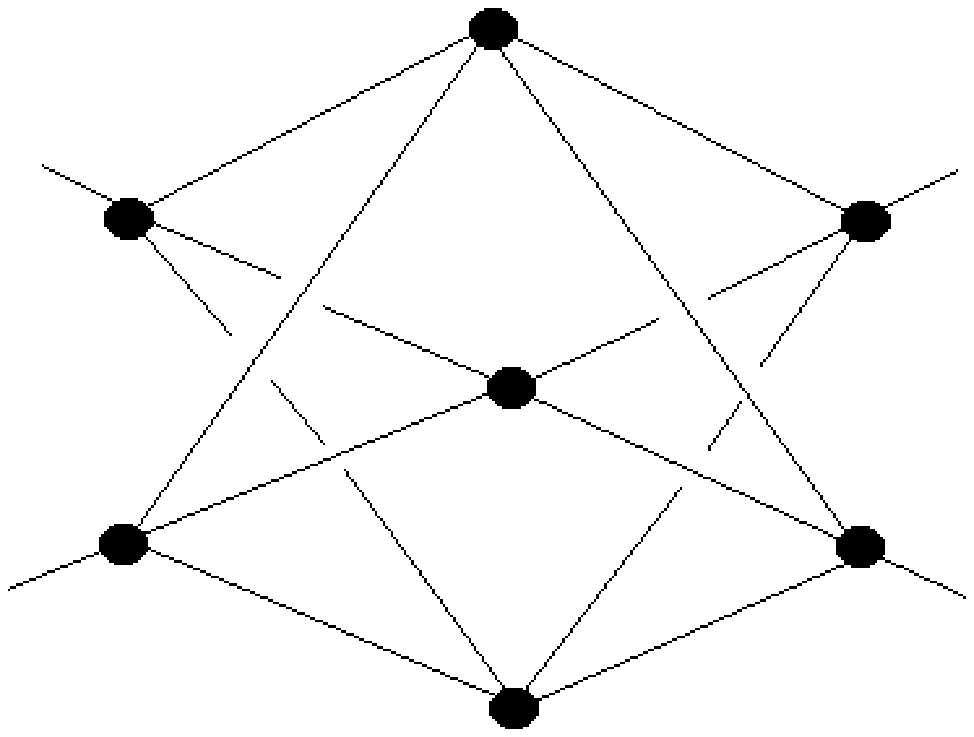}
\caption{A 6-loop graph for the coupling of $\phi^4$ theory,
giving a non-zeta counterterm associated with the 8-crossing
positive 3-braid knot, $8_{19}$.}
\label{f1}
\end{figure}
\newpage

\begin{figure}[ht]
\epsfysize=6cm\epsffile{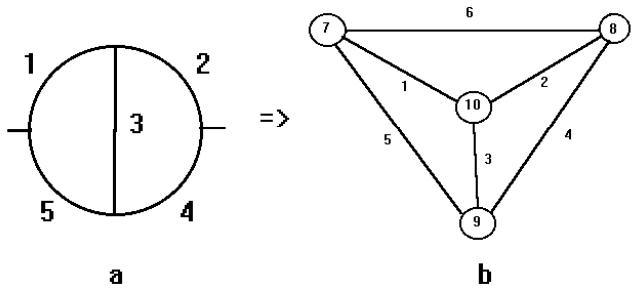}
\caption{The two-loop two-point function (a) is obtained by cutting
the log-divergent tetrahedral vacuum diagram (b) on the line with index
$\alpha_6=3d/2-\sum_{n=1}^5\alpha_n$.
The index $\alpha_{10}=\alpha_1+\alpha_2+\alpha_3-d/2$ is associated with
the vertex where lines 1,2,3 meet.}
\label{f2}
\end{figure}
\newpage

\begin{figure}[ht]
\epsfxsize=13cm\epsffile{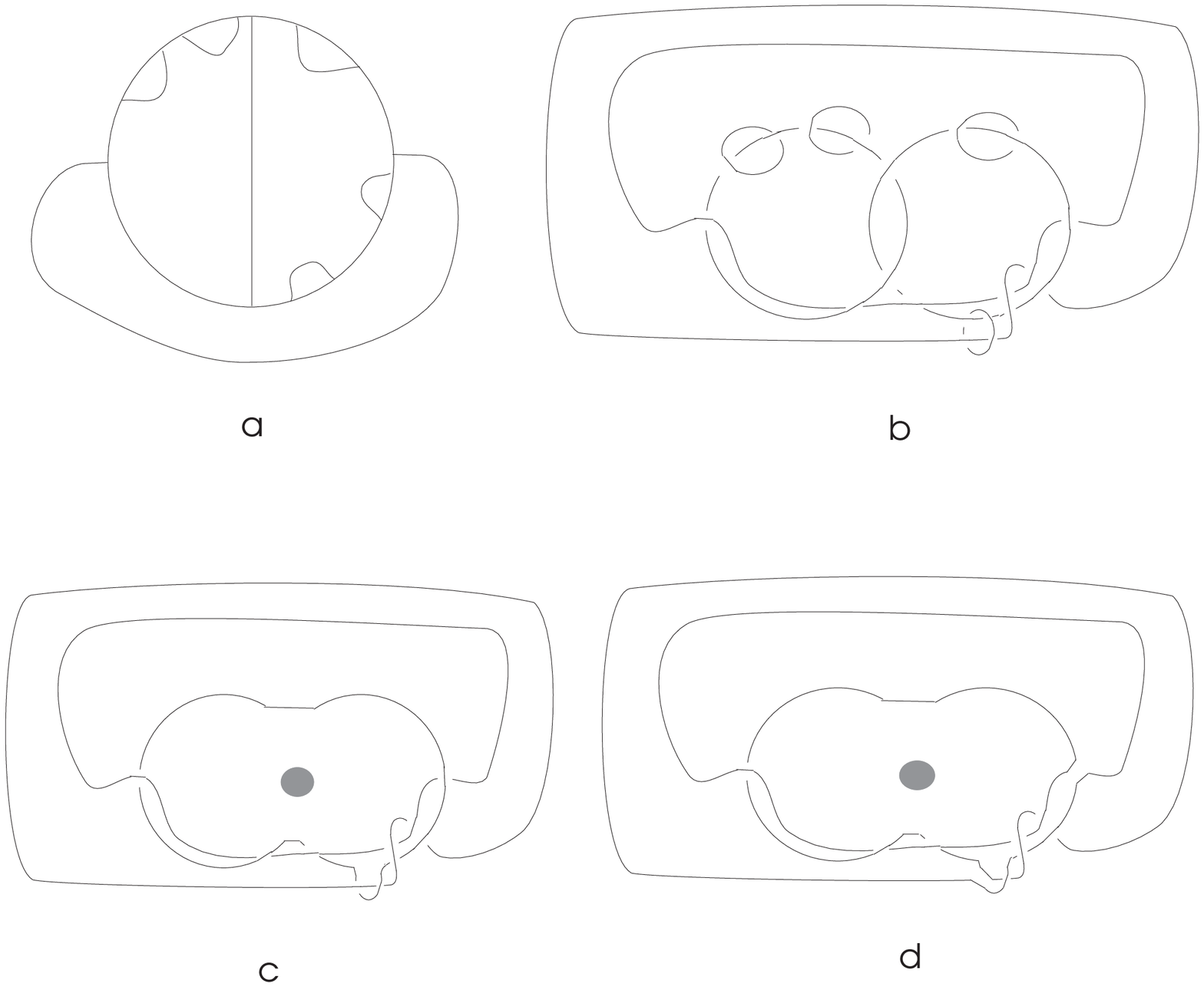}
\caption{Generation of $8_{19}=
\sigma_1^{}\sigma_2^3\sigma_1^{}\sigma_2^3$ by self-energy insertions.}
\label{f3}
\end{figure}
\newpage

\begin{figure}[ht]
\epsfxsize=13cm\epsffile{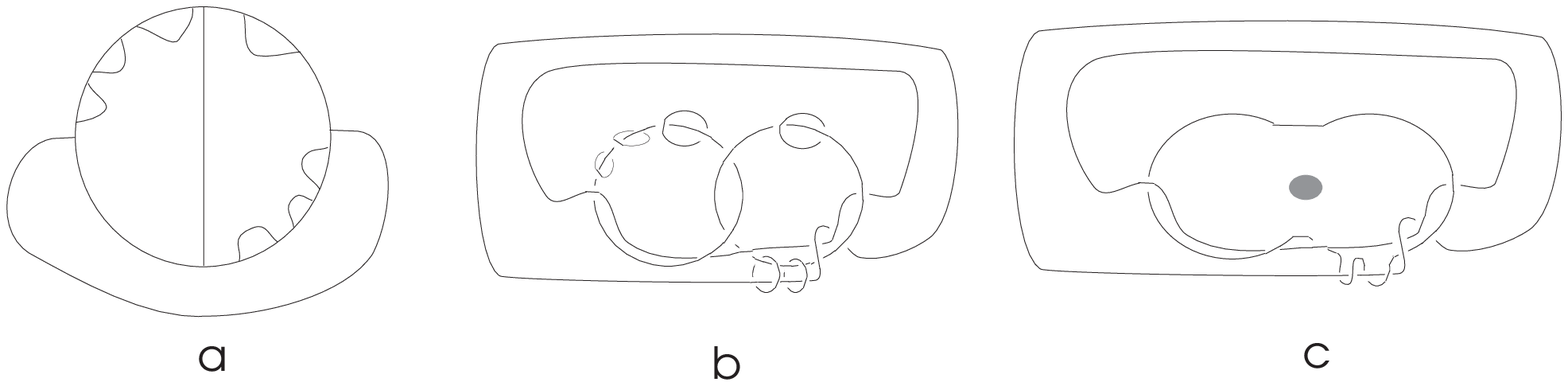}
\caption{Generation of $10_{124}=
\sigma_1^{}\sigma_2^5\sigma_1^{}\sigma_2^3$ by self-energy insertions.}
\label{f4}
\end{figure}
\newpage

\begin{figure}[ht]
\epsfxsize=13cm\epsffile{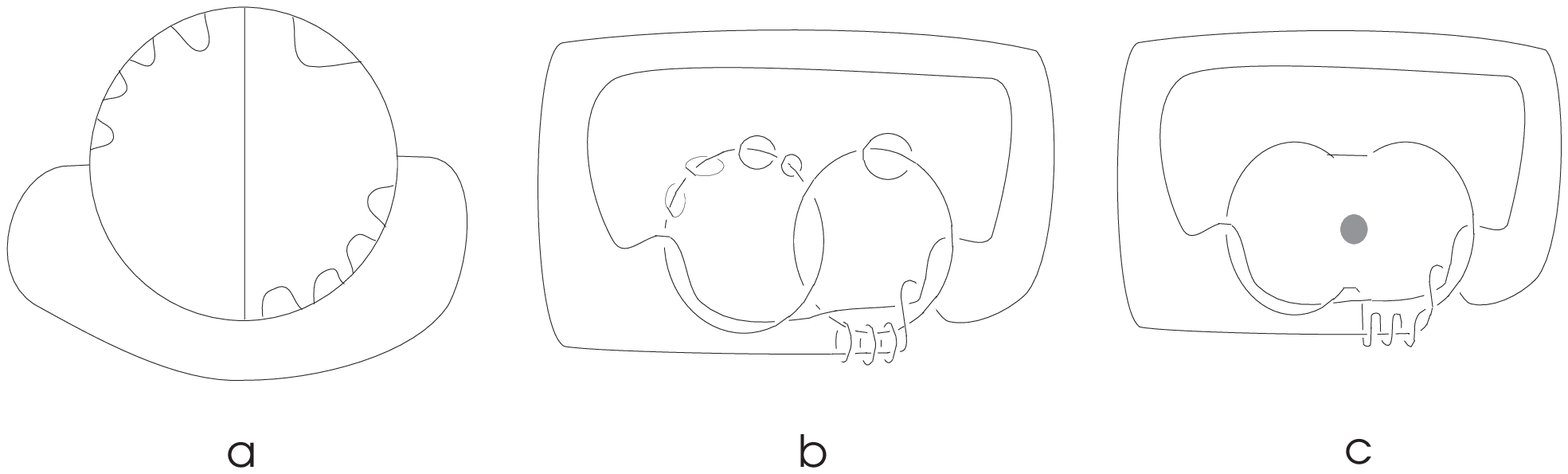}
\caption{Generation of the 12-crossing 3-braid
$\sigma_1^{}\sigma_2^7\sigma_1^{}\sigma_2^3$ by self-energy insertions.}
\label{f5}
\end{figure}
\newpage

\begin{figure}[ht]
\epsfxsize=13cm\epsffile{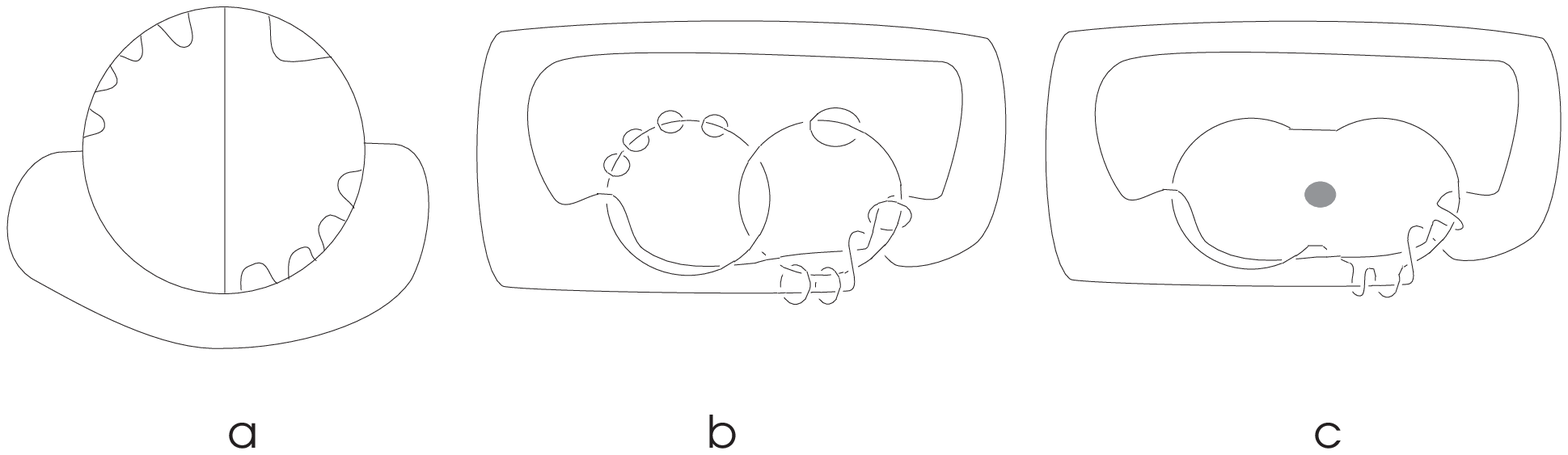}
\caption{Generation of the 12-crossing 3-braid
$\sigma_1^{}\sigma_2^5\sigma_1^{}\sigma_2^5$ by self-energy insertions.}
\label{f6}
\end{figure}
\newpage

\begin{figure}[ht]
\epsfxsize=13cm\epsffile{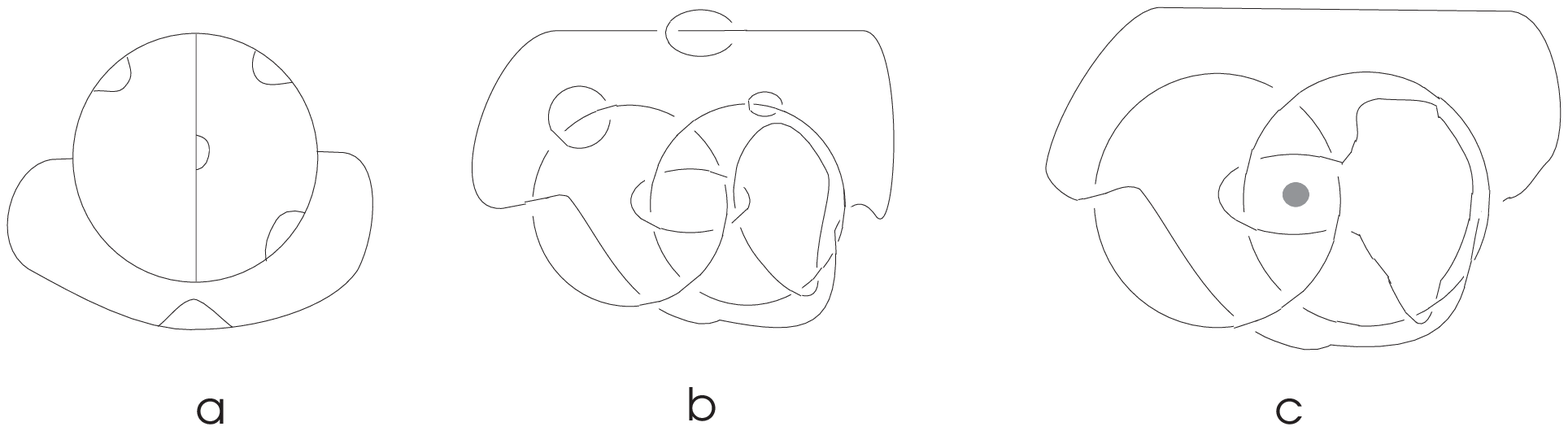}
\caption{Generation of the 9-crossing 4-braid positive factor knot
$\sigma_1^3\sigma_2^3\sigma_3^3$ by self-energy insertions.}
\label{f7}
\end{figure}
\newpage

\begin{figure}[ht]
\epsfxsize=13cm\epsffile{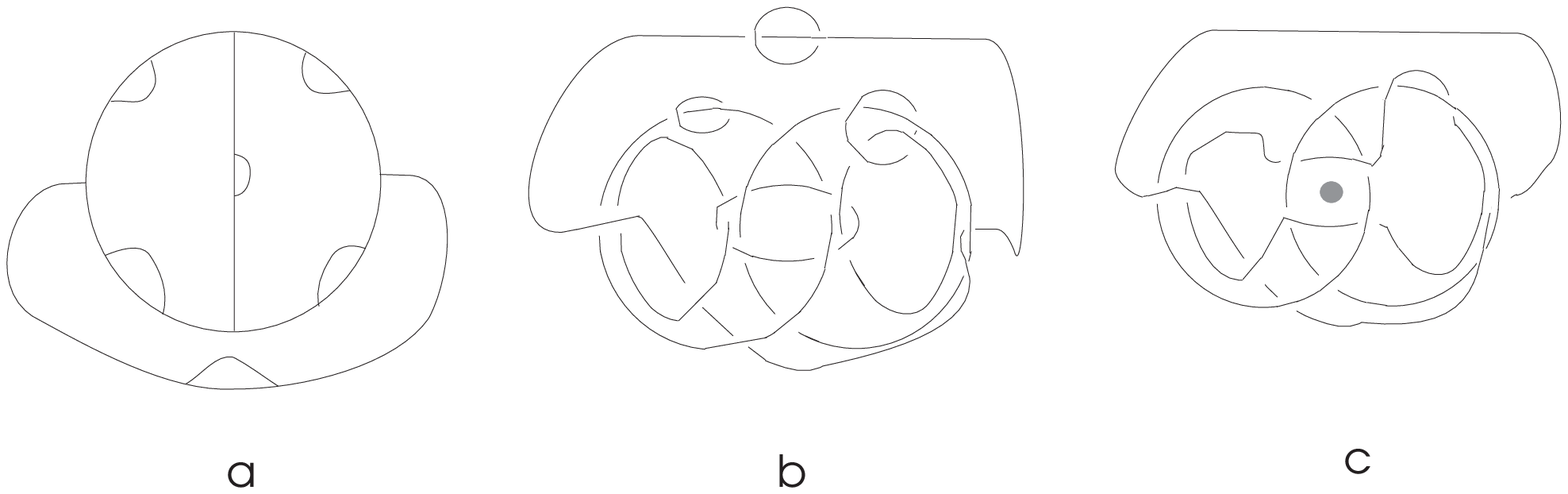}
\caption{Generation of the 11-crossing 4-braid positive prime knot
$\sigma_1^2\sigma_2^2\sigma_1^{}\sigma_3^{}\sigma_2^3\sigma_3^2$
by self-energy insertions.}
\label{f8}
\end{figure}

\end{document}